\documentclass[extra,mreferee]{gji}
\usepackage{timet}
\usepackage{graphicx} 
\usepackage{amsmath}
\usepackage{float}	 
\usepackage{subfigure}
\usepackage{icomma}  
\usepackage{multirow}
\usepackage{textcomp}
\usepackage{caption}
\usepackage{multicol}  
\usepackage{amssymb}
\usepackage[caption=false]{subfig}

\title[\texttt{Time-lapse reconstruction of the fracture front from diffracted waves arrivals} ]  
  {Time-lapse reconstruction of the fracture front from diffracted waves arrivals in laboratory hydraulic fracture experiments} 
  
\author[Dong Liu, Brice Lecampion, Thomas Blum]{Dong Liu $^1$, Brice Lecampion$^1$ \thanks{Corresponding author: brice.lecampion@epfl.ch}, Thomas Blum$^1$\\
$^{1}$Geo-Energy Laboratory- Gaznat chair on Geo-Energy, Ecole Polytechnique Fédérale de Lausanne,\\ ENAC-IIC-GEL-EPFL, Station 18, CH-1015, Switzerland}

\date{Received \today; in original form \today}
\pagerange{\pageref{firstpage}--\pageref{lastpage}}
\volume{?}
\pubyear{?}

\begin{document}
\label{firstpage}
\maketitle

\begin{summary}
4D acoustic imaging via an array of 32 sources / 32 receivers is used to monitor hydraulic fracture propagating in a 250~mm cubic specimen under a true-triaxial state of stress. We present a method based on the arrivals of diffracted waves to reconstruct the fracture geometry (and fluid front when distinct from the fracture front). Using Bayesian model selection, we rank different possible fracture geometries (radial, elliptical, tilted or not) and estimate model error. The imaging is repeated every 4 seconds and provide a quantitative measurement of the growth of these low velocity fractures. We test the proposed method on two experiments performed in two different rocks (marble and gabbro) under  experimental conditions characteristic respectively of the fluid lag-viscosity (marble) and toughness (gabbro) dominated hydraulic fracture propagation regimes. In both experiments, about 150 to 200 source-receiver combinations exhibit clear diffracted wave arrivals. The results of the inversion indicate a radial geometry evolving slightly into an ellipse towards the end of the experiment when the fractures feel the specimen boundaries. The estimated modelling error with all models is of the order of the wave arrival picking error. Posterior estimates indicate an uncertainty of the order of a millimeter on the fracture front location for a given acquisition sequence. The reconstructed fracture evolution from diffracted waves is shown to be consistent with the analysis of $90^{\circ}$ incidence transmitted waves across the growing  fracture.

\end{summary}


\section{Introduction}

Hydraulic fractures (HF) are a class of tensile fractures that propagate in a material
as a result of fluid pressurization \citep{detournay2016mechanics}. They
are encountered in a number of industrial applications such as oil and gas production, geothermal energy and block caving mining. HFs also propagate naturally in the form of magmatic dikes \citep{rivalta2015review}, or at glaciers bed due to the sudden release of surface melt water lakes \citep{TsRi10}.
 Investigation of the growth of such fluid-driven fractures under controlled conditions at the laboratory scale plays an important role in order to validate theoretical predictions.

Since the early work of \cite{HuWi57}, the measurement of the hydraulic fracture geometry has evolved from simple postmortem observations after the experiment to the use of continuous monitoring techniques during fracture growth. These developments have been slow and in most cases only post mortem observations are reported although sometimes via high resolution X-ray CT  \citep{LiJu16}. 
A photometry method based on the intensity drop of a back-light source as it passes through a dyed fracturing fluid has been successfully used to monitor hydraulic fractures in transparent materials \citep{Bunger06}. 
Such an optical technique has allowed to measure the evolution of both the fracture extent and the full field of fracture opening. Combined with particle image velocimetry, it also allows to measure the fluid velocity field  in the growing fracture \citep{ohubbert2018experimental}. These experimental techniques have provided invaluable data sets and insights into hydraulic fracture growth in transparent materials, such as PMMA, glass and hydrogel. They have notably helped in validating important theoretical predictions of hydraulic fracture mechanics \citep{bunger2008experimental,WuBu08,Bunger13,xing2017laboratory}.
However, by definition, these optical methods can not be used in non-transparent materials.
In rocks, acoustic emission (AE) monitoring is the main technique used to track the evolution of rupture 
 (\cite{LoBy77,zoback1977laboratory,Ishi01,stanchits2014onset,StBu15,GoNa15,stoeckhert2015fracture} to cite a few). AE events, however, do not provide a direct measurement of fracture geometry as they are mostly associated with micro-slips around the growing fracture \citep{RoSt16}. The observations of 
 self-potential \citep{MoGl07,HaRe13} 
 during HF experiments correlate with pressure evolution 
 and appear to highlight fluid flow patterns but do not provide an accurate measurement of the growing fracture. 
 Advanced imaging techniques such as Neutron imaging have been recently attempted \citep{RoMa18} as well as  2D digital image correlation (DIC) \citep{JeKe15,zhao20}.
 Neutron imaging necessitates the use of relatively small specimen (to achieve sufficient resolution) while the application of 2D DIC imposes the use of intricate specimen geometry and boundary conditions not suited to hydraulic fracturing. 
 
 We focus here on active acoustic imaging, a method 
 akin to a 4D seismic survey at the laboratory scale in the ultrasonic range. Earlier studies \citep{MeMa84,de1996physical,Glas98,van1999influence,groenenboom2000monitoring} have shown its capability to obtain quantitative information during laboratory hydraulic fracture experiments. 
The wave-field interacts in different ways with the growing fracture. It can be diffracted by the fracture tip (as well as the fluid front if a lag is present near the fracture tip) but also reflected by and transmitted through the fluid-filled fracture. 
The evolution of transmitted waves has notably allowed to identify a dry region near the fracture tip (fluid lag) \citep{MeMa84,de1996physical}.
Records of the arrival times of waves diffracted by the fracture have enabled to estimate the evolving fracture tip position under the hypothesis of a horizontal radial fracture centered on the injection point \citep{groenenboom2000monitoring,groenenboom2000scattering}.
Opening of the fracture results in attenuation and delay of transmitted waves. It thus allows to evaluate the fluid layer thickness by matching the spectrum of the transmitted signals with the transmission coefficient of a three layers model \citep{groenenboom1998monitoring}. These two techniques (diffraction and transmission)
 have been shown to provide results in agreement with optical methods  \citep{kovalyshen2014comparison}.

In this paper, we improve the imaging of a growing hydraulic fracture by using an unprecedented amount of piezoelectric source/receiver pairs and relaxing the assumption of a centered horizontal radial fracture. 
We test the method on two experiments performed in quasi-brittle rocks (marble and gabbro) which may exhibit different fracture growth behavior compared to PMMA, plaster or cement-based materials used in the previous studies cited above.
We first present our experimental set-up, the rocks and experimental conditions used. After illustrating the type of diffracted waves measured in these experiments, we develop an inverse problem for the fracture / fluid front  reconstruction. This inversion is then performed repeatedly in time for each acquisition sequence.
 We use different shapes (ellipse, circle with or without tilt) to parameterize the fracture front geometry and use Bayesian model selection to rank these different models. 
We finally compare the results obtained from the inversion of diffracted waves with transmitted waves data.

\section{Experimental methods}
\subsection{Experimental set-up and specimen preparation}

\begin{figure}
\centering 
\includegraphics[width=0.8\linewidth]{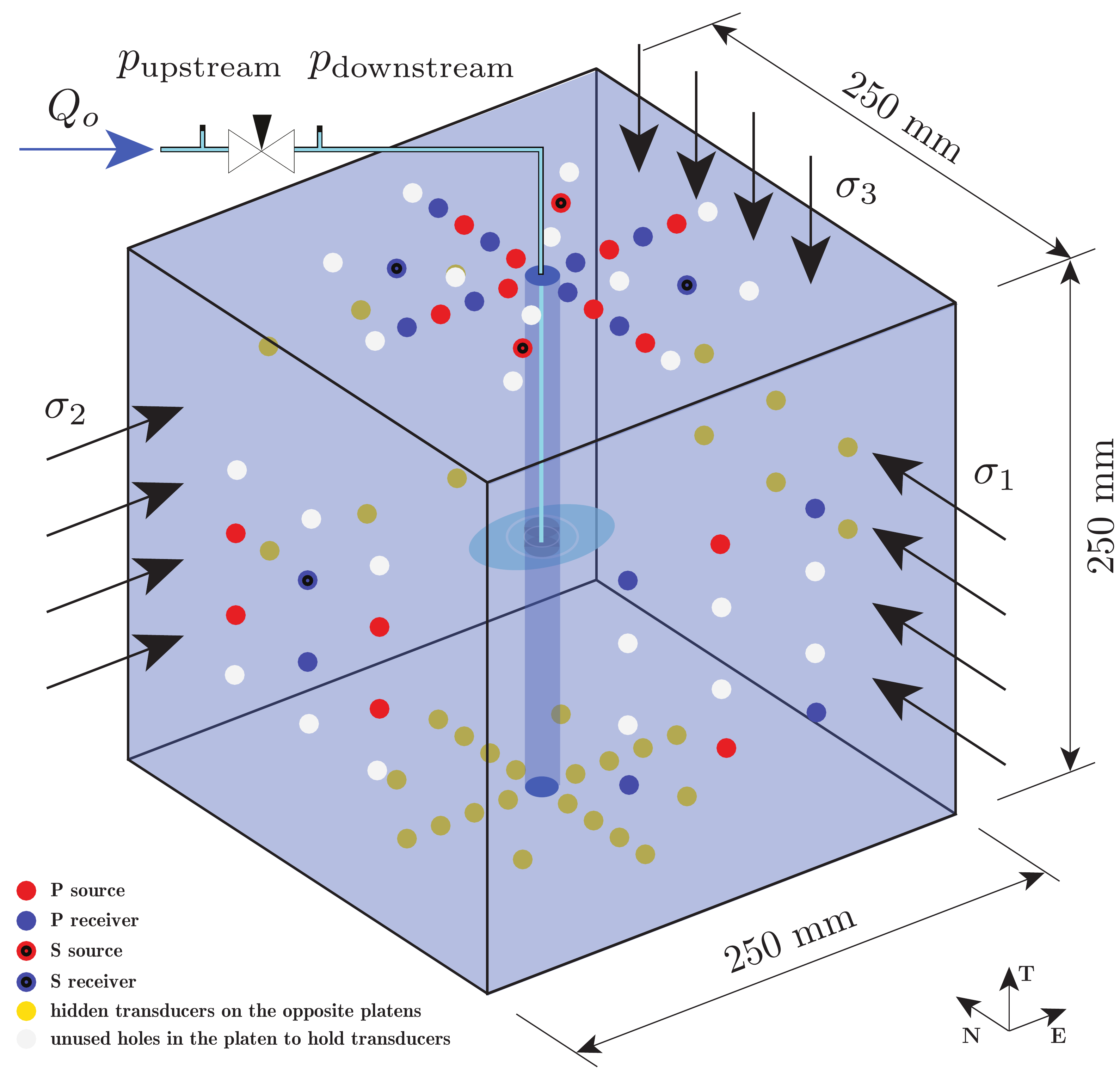} 
\caption{Schematic illustration of the rock sample showing the transducers disposition for the GABB-001 experiment. Additional holes are available in the platens allowing the use of various transducer dispositions. Two facing platens share the same transducers disposition and  source / receivers transducers are alternately located on opposite platens for robustness.
}
\label{fig:Transducers}
\end{figure}

Hydraulic fracture growth experiments are carried out in 250~mm cubic rock samples under a true triaxial compressive state of stress as illustrated in Fig.~\ref{fig:Transducers}. 
The confinement is applied by symmetric pairs of flat jacks in the three axis of a poly-axial reacting frame. Compressive stresses up to 20 MPa can be applied prior to injection. 
The fracturing fluid is injected in a central wellbore by a syringe pump (ISCO D160) at a constant flow rate (in the range 0.001~mL/min to 107~mL/min) with a maximum injection pressure of 51~MPa. An interface vessel in the injection line allows to inject a wide range of fluid type with viscosity ranging from 1 mPa.s to 1000 Pa.s.
Due to the compliance of the injection system (i.e. volume of fluid in the injection line), upon fracture initiation the flow rate entering the fracture does not equal the pump injection rate $Q_o$ during a transient phase \citep{lhomme05,LeDe17}.  A needle valve is thus placed in the injection line close to the well-head in order to control the release of fluid compressed during the pressurization phase. Using volume conservation within the injection system (pump to fracture inlet), it is possible to estimate the flow rate $Q_{in}(t)$ entering the fracture by taking the derivatives of the fluid pressure measurements (see appendix \ref{app:inletflux} for details).

The rock sample is first rectified as a cube of 250~mm~$\times$~250~mm~$\times$~250~mm dimensions. We polish the specimen surfaces to minimize friction and to ensure a good contact between the piezo-electric transducers and the rock. 
A wellbore of 16 mm diameter is drilled through the block and a horizontal axisymmetric notch (with a diameter of  21~mm $\pm$~1~mm) is created in the middle of the sample via a specifically designed rotating cutting tool. The resulting notch is axisymmetric and its plane perpendicular to the well axis.
A completion tool connected to a injection tubing is epoxied in the wellbore and allows to inject fluid only at the notch level.

Active acoustic monitoring is integrated within the poly-axial cell. 64 piezoelectric transducers are included in the loading platens: 32 transducers act as sources and 32 as receivers.  This array of transducers consists of 10 shear-wave transducers and 54 longitudinal-wave ones.
We use a source function generator connected to a high-power amplifier to send a Ricker excitation signal with a given central frequency that can be set between 300 and 750~kHz depending on the material type. The source signal is routed to one of the 32 source transducers via a multiplexer. The 32 receiver transducers are connected to a high-speed acquisition board in order to record the signal simultaneously on all receivers with a sampling frequency of 50~MHz.
As the switch between sources is limited by the multiplexer, the excitation of a given source is repeated 50 times and the data stacked to improve signal to noise. Spanning of the 32 sources defines an acquisition sequence and takes about 2.5 seconds in total. 
In addition to acoustic data, we record fluid injection pressure (upstream and downstream the needle valve), volume and pressure of each flat-jack pairs at 1 Hz. All the measurements are synchronized via a dedicated LabView application.

\begin{table}
\centering
\caption{Material properties ($V_p$ and $V_s$ are measured directly on the cubic rocks 
during the pressurization stage prior to any fracture growth). 
\label{tab:Materials}} 
\begin{tabular}{c|c|c|c|c}
\hline
Rock & $V_p$ (m/s) & $V_s$ (m/s) & $\rho$ ($\times 10^3$kg/m$^3$) & Grain size (mm) \\
\hline
Carrara marble & 6249.8 $\pm$ 54.0 & 3229.9 $\pm$ 176.2 & 2.69 & 0.1-0.2 \\
Zimbabwe gabbro  & 6679.0 $\pm$ 113.2 & 3668.5 $\pm$ 41.3 & 3.00 & 1-3 \\
\hline
\end{tabular}
\end{table}

\begin{table}
\centering
\caption{Sample configuration and experimental parameters for Zimbabwe gabbro and Carrara marble.
 \label{tab:ExpInfo}} 
\begin{tabular}{c|c|c|c|c}
\hline
Rock & Block size &  $\sigma_3$ & $\sigma_1=\sigma_2$   & Location of the notch from \\
sample & (mm) & (MPa) & (MPa) & the sample bottom $x_3$ (mm)\\
\hline
GABB-001 & $250\times 250\times 251$ &  0.5 & 10.5 & 128.5\\
MARB-005 & $257\times 256\times 256$ & 10 & 20 & 131\\
\hline
& Fracturing fluid & Viscosity & Injection rate & System compliance\\
&  & $\mu$ (Pa.s) &$Q_o$ (mL/min) & $U$ (mL/GPa)\\
\hline
GABB-001 &  Glycerol & 0.6 & 0.2 & 217.3 \\
MARB-005 &  Silicone oil & 100 & 0.2 & 282.5 \\
\hline
\end{tabular}
\end{table}

\begin{table}
\centering
\caption{Characteristic time-scales for a radial hydraulic fracture: viscosity-toughness transition $t_{mk}$, fluid lag disappearance time scale $t_{om}$. 
We estimate these time-scales using the averaged entering flow rate into the fracture $<Q_o>$ during the fracture duration $t_{prop}$ and the
following material properties: a) for Zimbabwe gabbro, $E=68.4$~GPa \citep{TruStone}, $\nu=0.3$ (assumed), $K_{Ic}=3.03$ MPa.m$^{1/2}$ \citep{meredith1985fracture}; b) for Carrara marble, $E=65$~GPa, $\nu=0.25$ \citep{gulli2015mechanical}, $K_{Ic}=1.38$ MPa.m$^{1/2}$ \citep{ouchterlony1990fracture}. 
\label{tab:ExpScaling}}

\begin{tabular}{c|c|c|c|c}
\hline
Rock & $<Q_o>$  & Propagation  & $t_{mk}$  & $t_{om}$ \\
& (mL/min) & duration $t_{prop}$ (s) & (s) & (s) \\ 
\hline
    GABB-001   & 0.074 & $\approx$ 410 & $4.0\times 10^{-4}$  & $3.3\times 10^5$ \\
MARB-005  & 0.046 & $\approx$ 582 & $5.0\times 10^4$ & $5.8\times 10^3$ \\
\hline
\end{tabular}
\end{table}

\subsection{Laboratory hydraulic fracturing experiments}

\begin{figure}
\centering 
\includegraphics[width=0.8\linewidth]{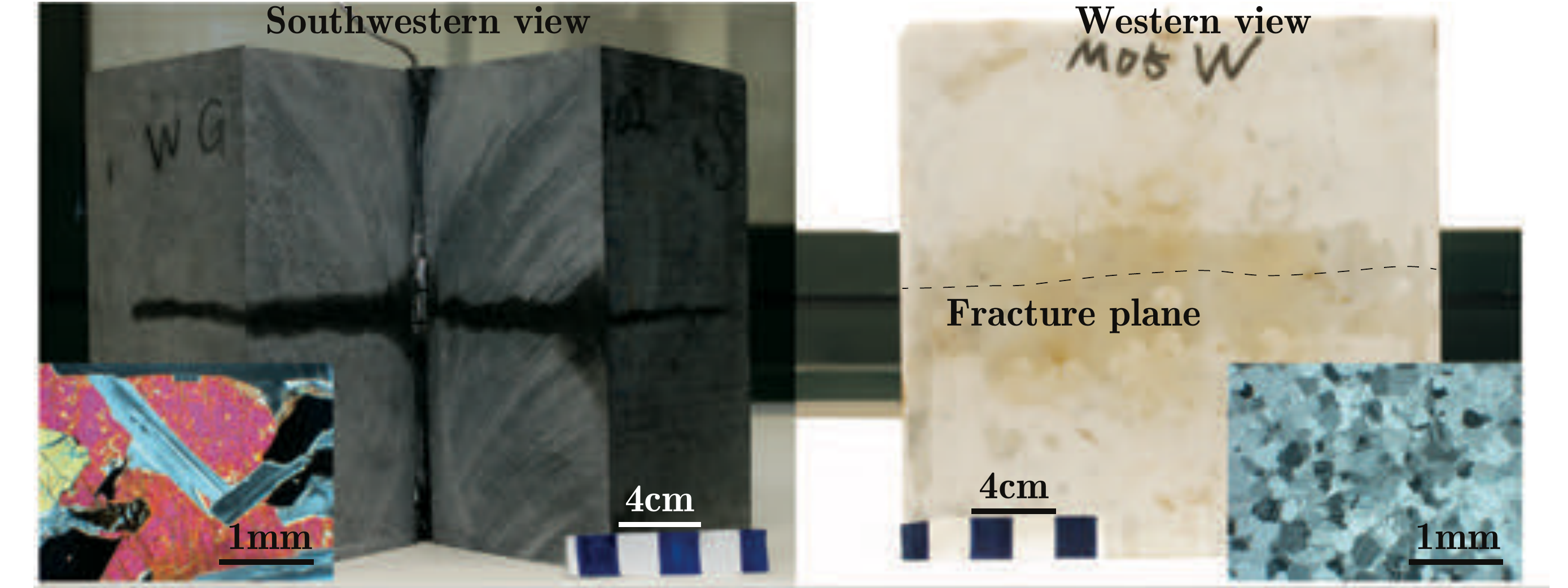} 
\caption{Thin section of Zimbabwe gabbro and Carrara marble and the postmortem photos of the cut blocks after GABB-001 and MARB-005 experiments. The wet region around the fracture in the gabbro block does not necessarily indicate a large leak-off during the injection. The visible inhibition occurred mostly after the experiment as the block was cut and photographed 25 days after the experiment.}
\label{fig:PostMortem}
\end{figure}

We discuss in this paper two experiments performed respectively in Zimbabwe gabbro and Carrara marble. Zimbabwe gabbro (plagioclase, mica, biotite and amphibole, quartz) has a larger grain size than Carrara marble (calcite, mica) as illustrated in Fig.~\ref{fig:PostMortem}, which 
usually implies a larger fracture process zone \citep{ouchterlony1982review}. Both rocks have isotropic acoustic properties (see in Table.~\ref{tab:Materials}). 
We impose a bi-axial state of stress on the block setting $\sigma_1=\sigma_2$, while the 
the minimum stress $\sigma_3<\sigma_1=\sigma_2$ is set perpendicular to the wellbore in order to favour a planar fracture initiating from the axisymmetric notch, in other words, promoting a transverse fracture to the wellbore (as shown in Fig.~\ref{fig:PostMortem}). Glycerol (gabbro) or silicone oil (marble) are used as fracturing fluids. The fluid is injected at a constant rate. All experimental parameters are listed in Table \ref{tab:ExpInfo}. 
The active acoustic monitoring is conducted with a central frequency of 750~kHz for the source excitation. This results in a wavelength of around 9~mm for compressional waves in both rocks whose grain size is at most 3~mm in gabbro and 0.2~mm in marble. 
In both experiments, acoustic acquisition is performed with a larger period during the pressurization phase and then switched to every 4 seconds during fracture propagation as shown in Fig.~\ref{fig:G01NonAcoustic} and Fig.~\ref{fig:M05NonAcoustic}.

\subsubsection{Injection Design}
The propagation of a fluid-driven fracture is a multiscale physical process. 
For a radial hydraulic fracture propagating in a tight material, it is now well established that initially at early time due to the injection of a viscous fluid, a fluid-less cavity (fluid lag) forms at the fracture tip (see \cite{detournay2016mechanics} and references therein). 
The fluid front then catches up with the fracture front over a characteristic time-scale
\begin{equation}
t_{om}=E^{\prime 2} \mu^\prime /\sigma_3^3
\end{equation}
where $E^\prime=\frac{E}{1-\nu^2}$ is the plane-strain elastic modulus related to the Young's modulus and Poisson's ratio, $\mu^\prime=12\mu$ with $\mu$ the fluid viscosity and $\sigma_3$  is the minimum applied stress (normal to the fracture plane).
In addition, as the perimeter of the radial fracture grows, the energy spent in creating new fracture surfaces increases.
The propagation switches from a regime dominated by fluid viscosity to a regime dominated by fracture toughness. This evolution is captured by a dimensionless toughness $\mathcal{K}=(t/t_{mk})^{1/9}$, where $t_{mk}$ is the transition time-scale from the viscosity to toughness dominated regimes of growth:
\begin{equation}
    t_{mk}=\frac{E^{\prime 13/2}Q_o^{3/2}\mu^{\prime 5/2}}{K^{\prime 9}},\quad K^\prime=\sqrt{\frac{32}{\pi}}K_{Ic},
\end{equation}
where $K_{Ic}$ is the mode I fracture toughness. 
More precisely, the fracture grows in the viscosity dominated regime as long as $\mathcal{K}\lesssim 1$ and strictly in the toughness dominated regime for $\mathcal{K}\gtrsim 3.5$ \citep{savitski2002propagation}. Moreover, the fluid lag vanishes at all time in the toughness dominated regime \citep{lecampion2007implicit}. In other words,  if $t_{mk}\ll t_{om}$, no fluid lag is observed. 

Experimentally, for a given rock, one can adjust the injection rate $Q_o$, fluid type (viscosity $\mu$) and the minimum stress perpendicular to the fracture plane $\sigma_3$ to explore a given propagation regime. 
We refer to \cite{BuJe05} 
for the proper scaling and experimental design of laboratory hydraulic fracture experiments.
The two experiments reported herein are characteristic examples of two very different hydraulic fracture propagation regimes (toughness and lag-viscosity dominated). Table \ref{tab:ExpScaling} lists the corresponding time-scales estimated using 
values of the rock properties from the literature and the averaged injection rate.  In both experiments, the compliance of the injection system is rather important. As a result, the flow rate entering the fracture $Q_{in}$ is not constant and equal to the pump rate. It can however be estimated from the injection pressure and injection system compliance (see appendix \ref{app:inletflux} for details).
The GABB-001 experiment is toughness dominated as the propagation time is much larger than the viscosity - toughness transition time-scale and no fluid lag is expected during the fracture propagation ($t_{mk}<t_{om}$). 
On the other hand, the MARB-005 experiment is such that the propagation occurs in the so-called lag / viscosity dominated regimes \citep{LeDe17}: the propagation duration is smaller than both $t_{om}$ and $t_{mk}>t_{om}$.

\subsubsection{Toughness-dominated experiment GABB-001}
\begin{figure}
\centering 
\includegraphics[height=0.28\linewidth]{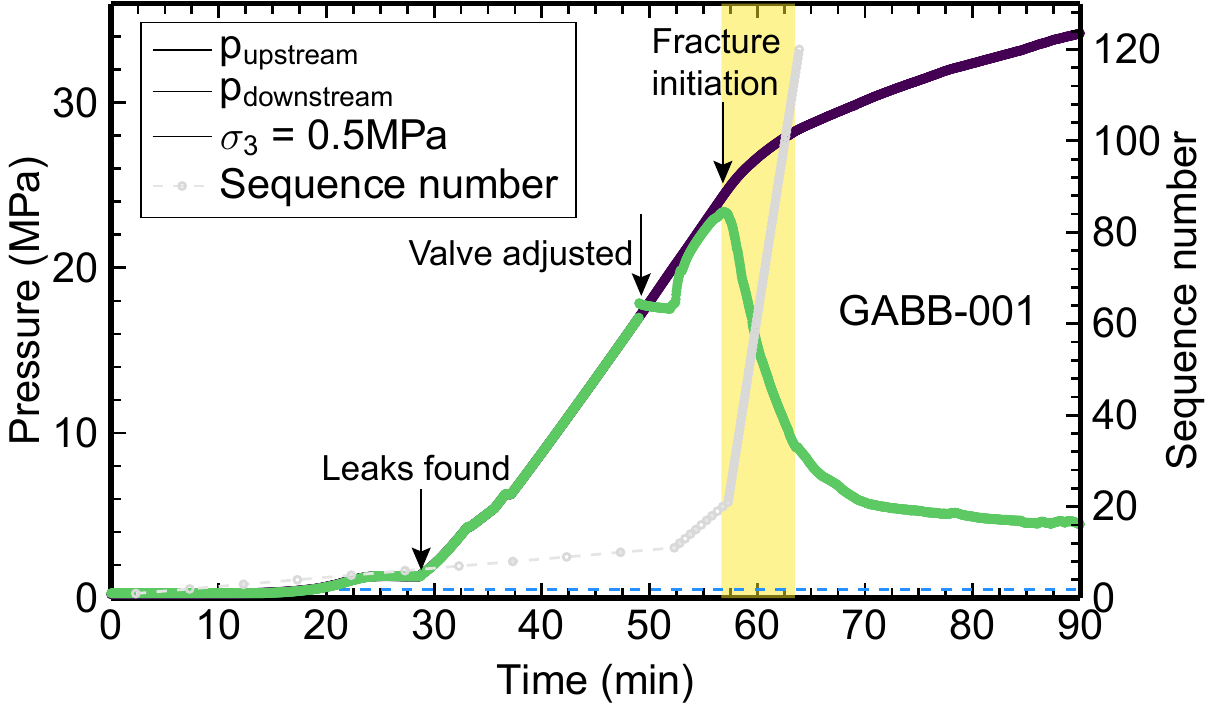}
\includegraphics[height=0.28\linewidth]{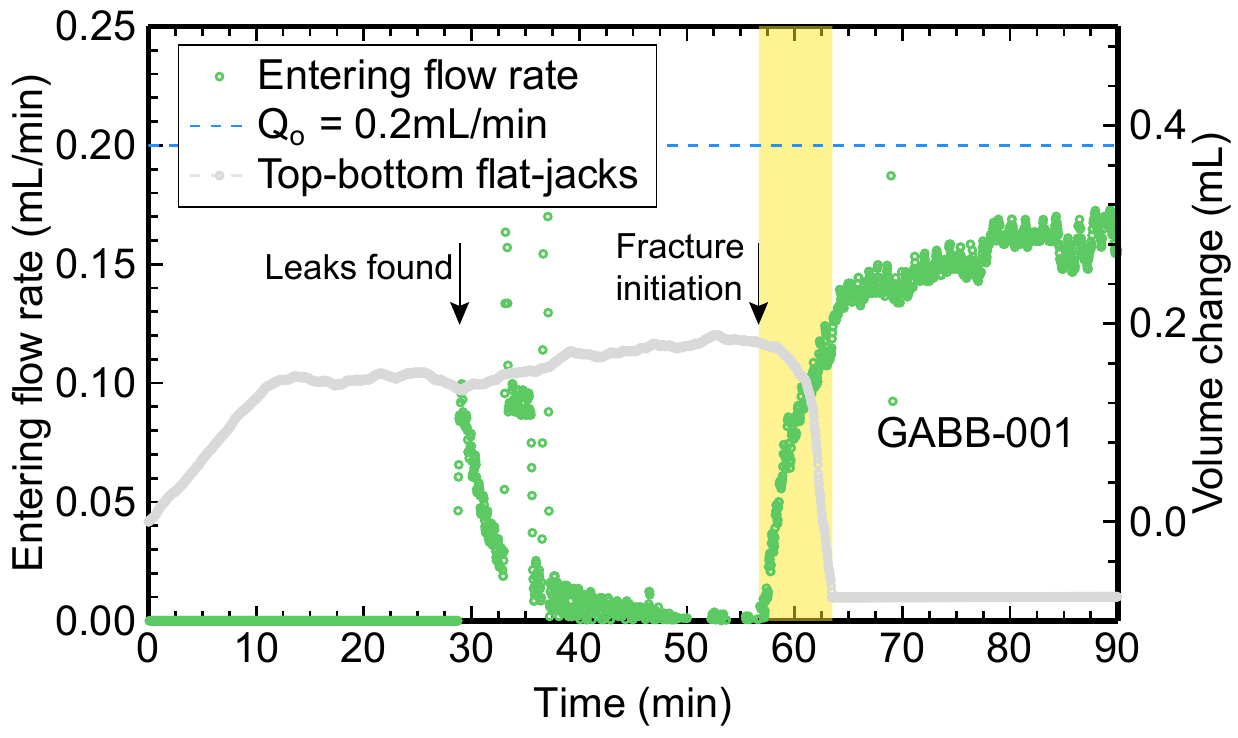}
\caption{GABB-001 experiment: evolution of the upstream and downstream pressure, and the mapping relation between the sequence number and acquisition time (left); and evolution of the entering flow rate of the fluid into the fracture, and the volume change of the top-bottom flat-jacks (right).  The yellow coloured time interval indicates the propagation of the fracture through the block (from the notch to the end of the block).}
\label{fig:G01NonAcoustic}
\end{figure}

The gabbro experiment (GABB-001) is a so-called toughness dominated experiment. As a result, the fluid pressure downstream of the valve responds almost instantly to fracture growth. As illustrated in Fig.~\ref{fig:G01NonAcoustic}, the pressure increases linearly with a pressurization rate $Q_o/U$ (pending the fix of an initial leak in the injection line and the adjustment of the needle valve). Upon fracture initiation from the notch, the needle valve prevents a complete sudden release of the fluid pressurized in the line: the pressure downstream of the valve drops while the upstream one keeps increasing illustrating the damping of the entering flux by the needle valve. The fracture initiation time is confirmed by the response of the flat-jacks volume parallel to the fracture plane - as well as the acoustic data as later presented. When the fracture front reaches the edges of the block, there are no more constraints on its deformation and a sudden response of the flat-jacks is observed as well as a kink in the downstream injection pressure (see Fig.~\ref{fig:G01NonAcoustic}). 

\subsubsection{Lag-viscosity experiment MARB-005}
\begin{figure}
\centering 
\includegraphics[height=0.28\linewidth]{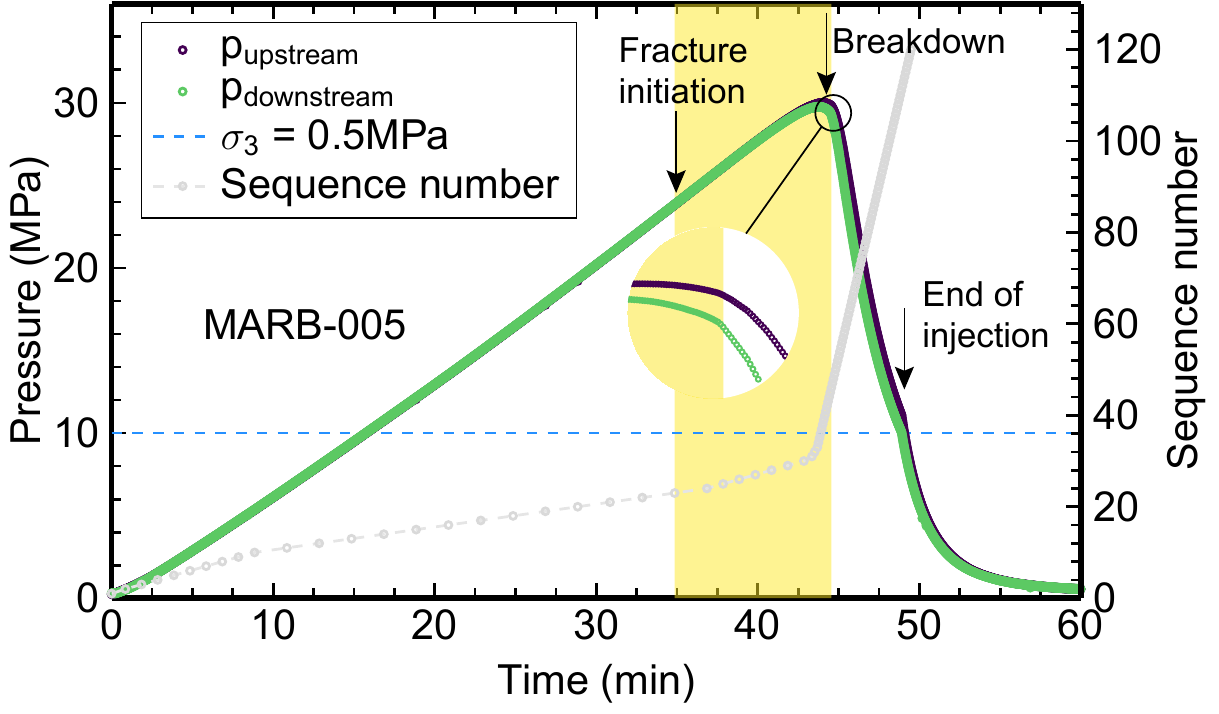}
\includegraphics[height=0.28\linewidth]{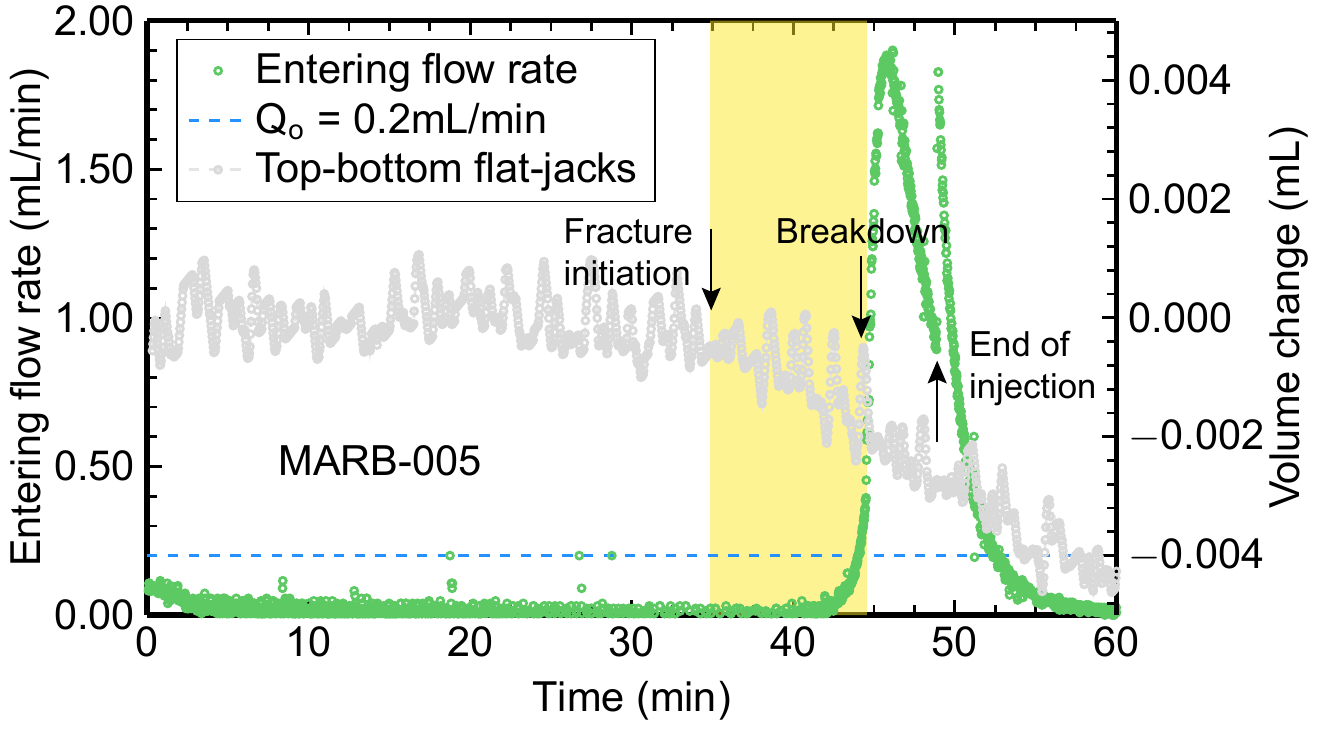}
\caption{MARB-005: evolution of the upstream and downstream pressure, and the mapping relation between the sequence number and acquisition time (left); and evolution of the entering flow rate of the fluid into the fracture, and the volume change of the top-bottom flat-jacks (right). The volume change of the top-bottom flat-jacks in MARB-005 is much smaller than for the GABB-001 experiment. This is most likely due to the presence of air bubbles inside the flat-jacks. The yellow coloured time interval indicates the propagation of the fracture through the block (from the notch to the end of the block).}
\label{fig:M05NonAcoustic}
\end{figure}

The marble experiment (MARB-005) is characterised by a large fluid lag and strong viscous drop. 
A restriction tube with a diameter of $1$~mm was placed in the injection line instead of the needle valve. Silicone oil was used as a fracturing fluid. Due to the large viscous effect, it takes time for the fluid to enter the fracture. 
The fracture initiation can be estimated from the response of the top-bottom flat-jacks and the appearance of acoustic diffraction. It is however indistinguishable from the pressure record - inline with previous observations \citep{ZhHa96,LeDe17}.
As can be seen from  Fig.~\ref{fig:M05NonAcoustic}, the entering flow rate remains limited  after fracture initiation. A large fluid lag develops behind the fracture front (as can be seen from the acoustic data on Fig.~\ref{fig:Diffraction}). The pressure in the injection line keeps increasing until the fracture front almost reaches the edges of the block. 
The fluid front continues to grow afterwards but the elastic deformation (and thus the hydro-mechanical coupling) is now different. This results in an  increase of the entering flow rate which correlates with a small kink in the downstream pressure. We thus use this point to approximate the time at which the fracture reaches the end of the block. It is worth noting that the maximum pressure (often misleadingly denoted as the breakdown pressure) occurs just before that time.

\section{Examples of acoustic diffraction data}

\begin{figure}
\centering 
\includegraphics[width=0.48\linewidth]{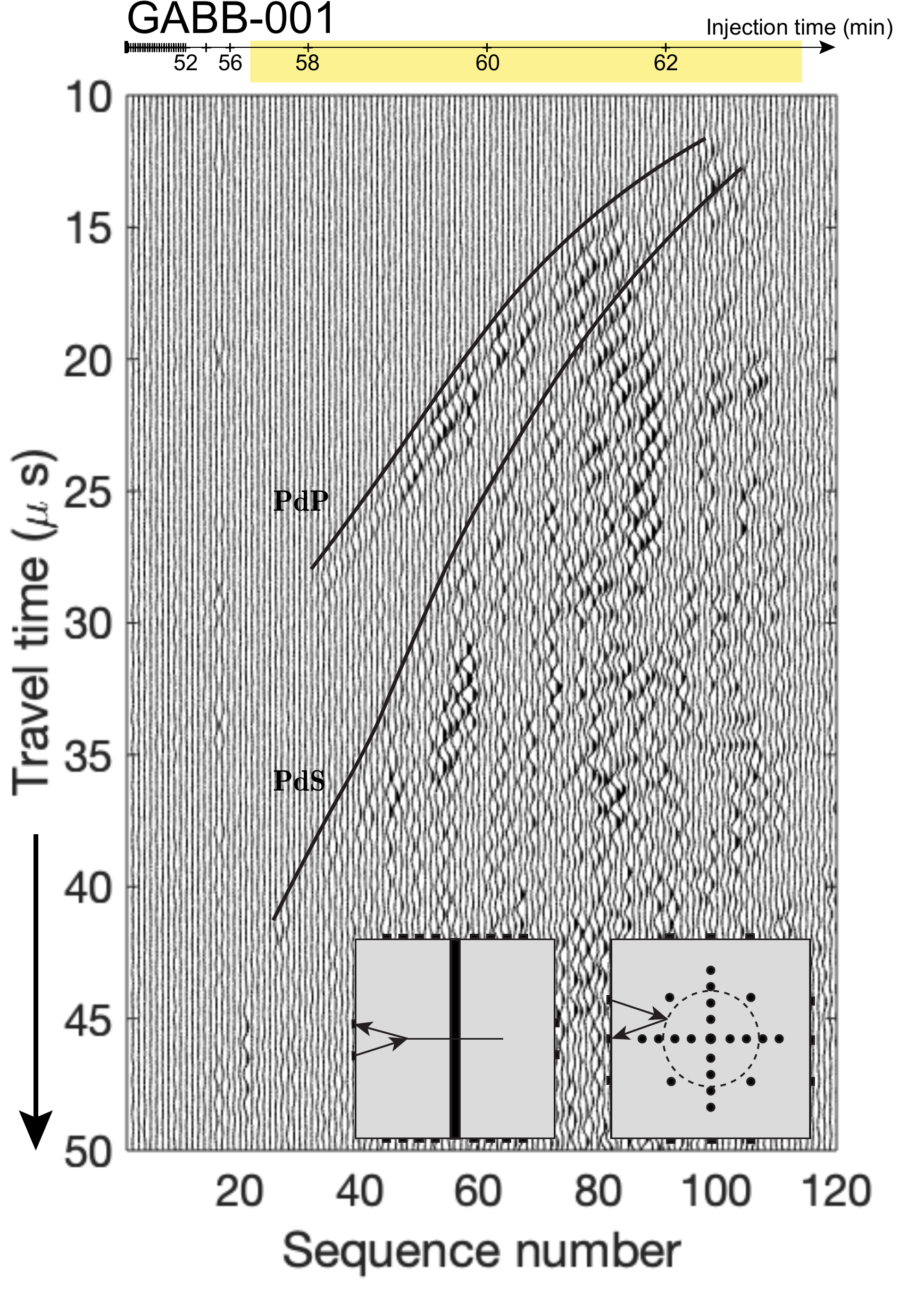}
\includegraphics[width=0.48\linewidth]{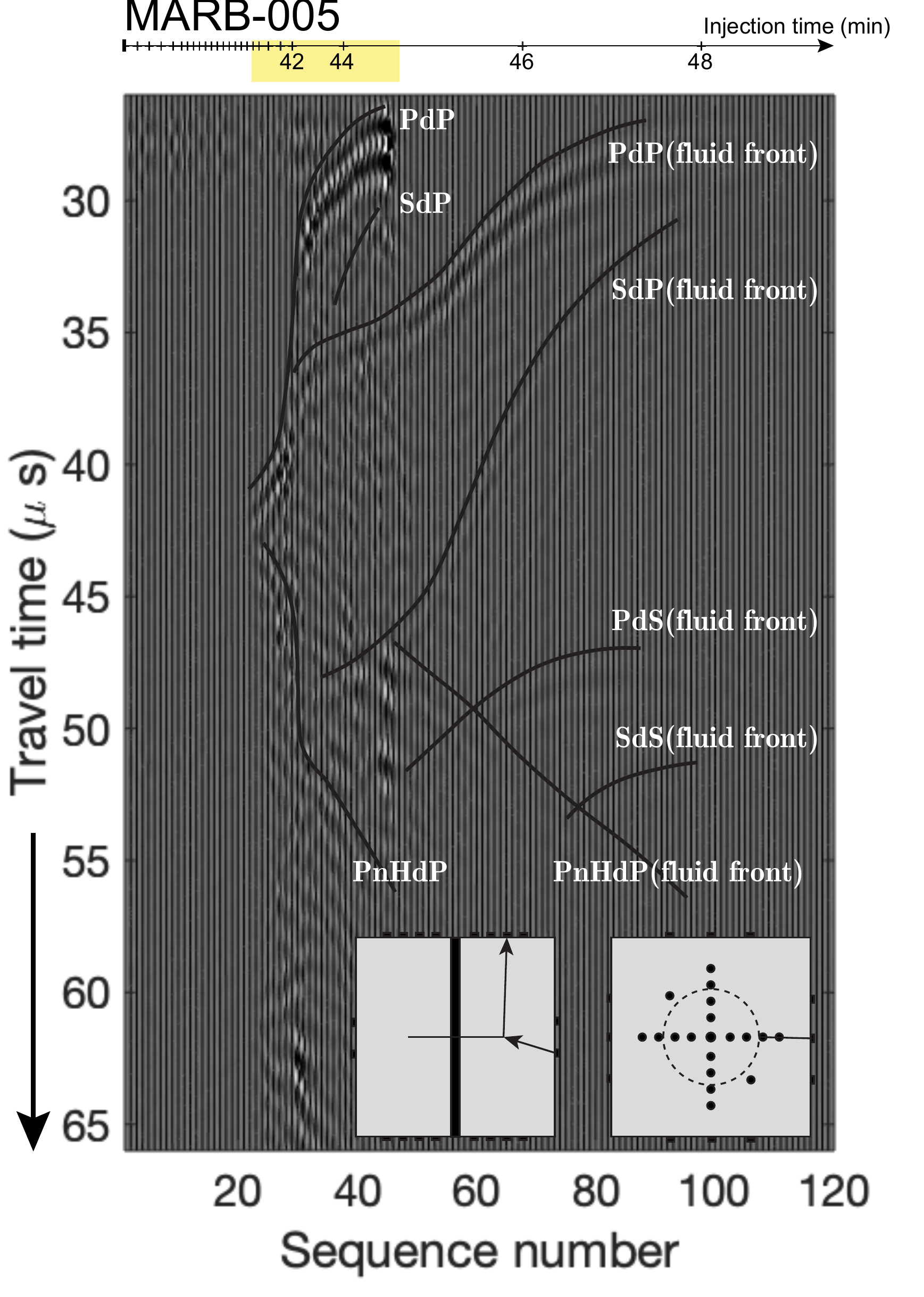}
\caption{Illustration of different diffracted waves pattern in GABB-001 and MARB-005 experiments. 
For MARB-005, a gray-scaled image of the data is shown combined with a wiggle plot. The yellow coloured time interval represents approximately the propagation of the fracture (from the notch to the end of the block). }
\label{fig:Diffraction}
\end{figure}
As observed in previous laboratory experiments \citep{groenenboom2000scattering,deGr01}, the initial notch, fracture front, and fluid front \citep{groenenboom2000monitoring} may all serve as a source of diffraction. Each receiver may record both compressional (P-wave) and shear (S-wave) components of the wave depending on its incident angle. Following \cite{groenenboom2000scattering},  
we categorise different acoustic events. We denote the diffraction along the fracture or fluid front with a 'd' and the interactions (reflection or diffraction) at the notch with an 'n'. 
We recall here some travel paths of diffracted waves 
observed in the two experiments presented here:
\begin{itemize}
    \item direct diffraction of the body wave at the fracture or fluid front with or without mode conversion: for example compressional waves diffracted by the fracture front without mode conversion (PdP), or shear waves diffracted by the fracture front with mode conversion (SdP).
    \item diffraction of the head wave (P wave guided by the fracture interface, denoted as H \citep{savic1995ultrasonic}) at the fracture tip with or without mode conversion. For example, PnHdP represents a P wave that is guided along the fracture interface after interacting with the notch. It is then diffracted at the fracture tip without mode conversion (see Fig.~\ref{fig:Diffraction}).  
\end{itemize}

Previous studies \citep{groenenboom2000scattering,deGr01} have observed more events related to reflections of the wavefield at the borehole tube and generalized Rayleigh waves propagating along the fracture. Such diffraction events arrive much later than the direct diffraction of body waves. 
Their signal to noise is often not sufficient to allow a proper picking of these later arrivals. We thus focus mainly on the diffracted body waves (PdP, PdS and SdP here).
As shown in Fig.~\ref{fig:Diffraction}, we are only able to observe PdP and PdS arrivals in GABB-001 for the chosen transducer pair, while for most transducer pairs only PdP arrivals can be clearly identified. In MARB-005, 
we are  able to recognize more diffracted waves, notably by both the fracture and the fluid fronts but also by the notch (PnHdP). In particular, the fluid front acts as a strong diffractor.

There are multiple techniques to visualize the evolution of the diffracted waves, as shown in Fig.~\ref{fig:Diffraction}. 
\cite{groenenboom1998acousticThesis,savic1995ultrasonic} have proposed to remove the direct incident wavefield by subtracting the signal recorded at a given time with the one recorded in the absence of the fracture at the beginning of the experiment.

This is sufficient to properly see the first diffracted wave arrival PdP. However,
we have found that for our experiments, subtracting 
the signal recorded at a given acquisition sequence with the one of the previous sequence provide a clearer image of the diffracted waves (notably the later ones). This is akin as a high pass filter along the experimental time axis (dimension of the sequence number) on the diffraction plots as in Fig.~\ref{fig:Diffraction}.

The reason why making the difference with the first sequence provides more blurry image is likely due to the evolution of the scattering background associated with the fracture roughness. The diffracted coda wave is much more noisy in the gabbro compared to the marble experiment, in line with the difference of rock grain size and posterior fracture roughness observed.

\section{Reconstruction of the fracture and fluid fronts using Bayesian inversion}

From the picking of the diffracted wave arrivals (for different source-receiver pairs), we invert for the geometry of the fracture or fluid front. We do so using different geometries for the diffraction front. We use a Bayesian framework to rank these different geometrical models.
The inverse problem is performed for each acquisition sequence independently in order to
finally obtain the time evolution of the fracture geometry.

\subsection{Forward models}
\begin{figure}
\centering 
\includegraphics[height=0.40\linewidth]{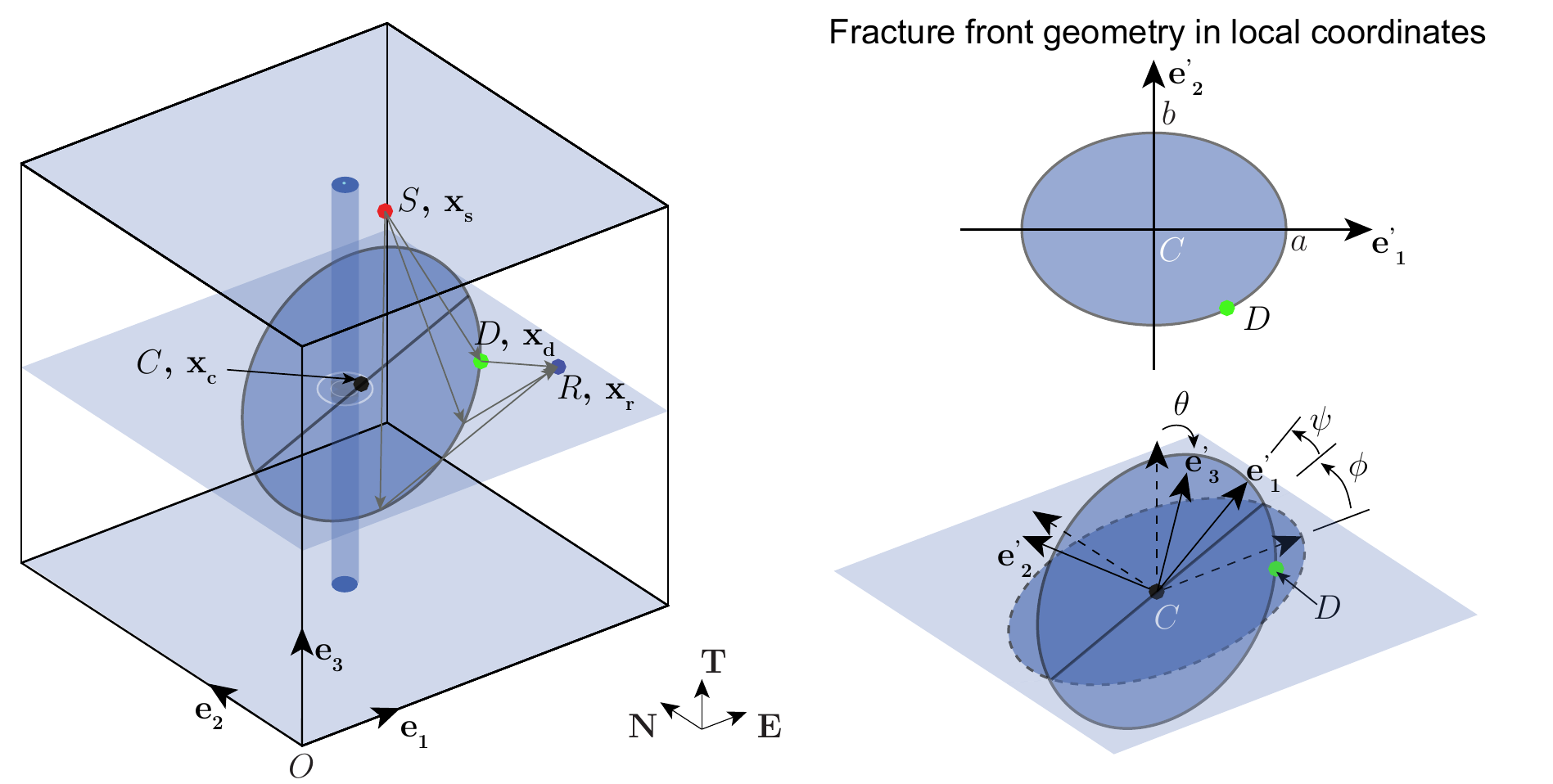}
\caption{Illustration of the fracture front geometry and its corresponding diffractor position for a given source-receiver pair. The diffractor position $D$ (in green) characterises the shortest travelling time between the source $S$ (in red) and the receiver $R$ (in blue) among all positions along a given fracture front. The fracture geometry is assumed to be planar and is defined in a local coordinate system ($\mathbf{e'_i}$, $i=1, 2, 3$) with its geometric center $C$ as the origin. The geometric description in the global coordinate system ($\bf{e_i}$, $i=1,2,3$) is  obtained after coordinates transformation (rotation and translation). }
\label{fig:DiffractionFrontIllustration}
\end{figure}

In the case of the direct diffraction of a body wave by the fracture front, 
the theoretical arrival $ t^d_{sr}$ of the diffracted wave for a chosen source-receiver pair is simply given by
\begin{equation}
      t^d_{sr} = \frac{\left\lVert \bf{x}_s-\bf{x}_d\right\rVert}{V_{sd}}+\frac{\left\lVert \bf{x}_d-\bf{x}_r\right\rVert}{V_{dr}}
    \label{eq:arrivaltime}
\end{equation}
where $V_{sd}$ and $V_{dr}$ are respectively velocities of the incident wave and diffracted wave (P-wave or S-wave). $\bf x_s$, $\bf x_r$ represent  the coordinates of the source and receiver transducers and $\bf x_d$  the coordinates of the diffractor, as illustrated in Fig.~\ref{fig:DiffractionFrontIllustration}. For a given fracture front geometry, $\bf x_d$ is obtained as the point which gives the minimal diffracted wave arrival $t_{sr}$ for a given source-receiver pair. We assume here that the diffraction front is planar. We parametrize it using a simple shape. We consider a possible
offset of the fracture geometric center $C$ with the injection point and denote $\mathbf{x_c}=(x_1, x_2, x_3)$ its coordinates. We also allow a possible tilt of the fracture plane captured by the three Euler angles: the dip $\theta$, azimuth $\phi$, and precession $\psi$ (see  Fig.~\ref{fig:DiffractionFrontIllustration}). 

We use an ellipse or a circle to describe the geometry of the diffraction front.
In order to account for a possible tilt of the fracture, we invert the data with four different forward models having different number of parameters $\mathbf{m}$ as listed in Table~\ref{tab:Model}. 
In the case of an elliptical diffraction front,  8 parameters describe its geometry:  the semi-lengths of the ellipse $a$ and $b$, $\mathbf{x_c}=(x_1, x_2, x_3)$ the position of the geometric center and three Euler angles $\phi, \theta, \psi$ characterising the fracture plane orientation (see Fig.~\ref{fig:DiffractionFrontIllustration}).

For a given geometrical model of the fracture front, we relate the measured diffracted arrivals for the different source-receiver pairs  with the forward predictions as 
\begin{equation}
{\bf d}=\mathbf{G}(\mathbf{m})+{\bf \epsilon}
\end{equation} 
where $\bf d$ denotes the picked arrival time for the different source-receiver pairs, $G(\mathbf{m})$ the arrival time predicted by the forward model, and ${\bf \epsilon}$ combines both measurement and modelling errors. For simplicity \citep{Tara05}, we will assume that $\epsilon$ follows a Gaussian distribution with zero mean and variance of $\sigma^2$. We will notably invert for $\sigma$ here thus providing a measure of the modelling error.

\begin{figure}
\centering 
\includegraphics[height=0.40\linewidth]{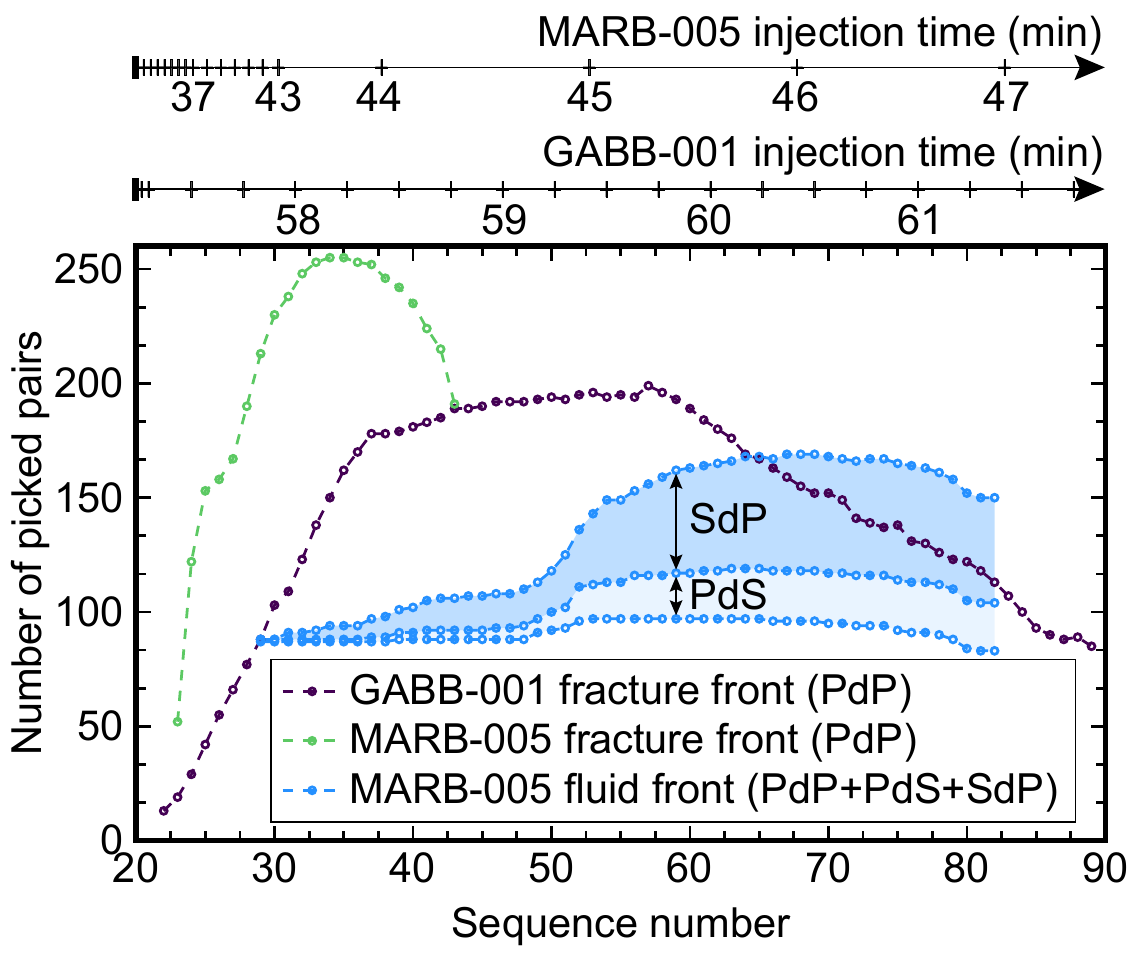}
\caption{The number of diffracted arrivals picked from different source-receiver pairs in GABB-001 and MARB-005 experiments. The corresponding acquisition time is indicated at the top of the figures with one tick every 15 seconds for GABB-001 and every minute for MARB-005.}
\label{fig:PickedPairsNumber}
\end{figure}

For a chosen source-receiver pair, we  manually pick the arrival time of the diffracted waves for different sequences using plots similar as Fig.~\ref{fig:Diffraction}. A spline is first drawn along the diffracted arrival. The coordinates of the spline passing through the corresponding sequence numbers are then collected as picked arrival time. 
It is worth noting that the number of picked source-receiver pairs may vary with time, as notably the diffracted arrivals are less visible for some pairs at early (close to initiation for small fractures) and late time (due to proximity of the fracture front with the edges of the block). 
The number of picked arrivals and their types of diffraction events for the different acquisition sequence is reported in Fig.~\ref{fig:PickedPairsNumber}.

\begin{table}
\centering
\caption{Model description and model parameters\label{tab:Model}. $N_p$ is the number of the parameters. $A=\ln a$, $B=\ln b$, $R=\ln r$ are adjusted fracture size parameters accounting for the values of original parameters, e.g. $a, b, r$, greater than zero. $\mathbf{x_c}=(x_1, x_2, x_3)$ characterises the offset of the geometric center of the diffraction front with respect to the origin of the global coordinate system.} 
\begin{tabular}{c|c|c|c|c}
\hline
Model  & Model description & $N_p$ & Model parameters $\bf m$  \\
\hline
$\mathcal{M}_1$ & Elliptical shape & 8 & $[A, B, x_1, x_2, x_3, \psi, \theta, \phi]$ \\
$\mathcal{M}_2$ & Circular shape ($a=b=r$, $\psi=0$) & 6 & $[R, x_1, x_2, x_3, \theta, \phi]$ \\
$\mathcal{M}_3$ & Horizontal elliptical shape ($\theta=\phi=0$) & 6 & $[A, B, x_1, x_2, x_3, \psi]$ \\
$\mathcal{M}_4$ & Horizontal circular shape ($\psi=\theta=\phi=0$) & 4 & $[R, x_1, x_2, x_3]$\\
\hline
\end{tabular}
\end{table}

\subsection{Inverse problem}
We seek to estimate both the model parameters $\bf m$ as well as the measurement/model error $\sigma$ \citep{LeGu07}. The likelihood of the data being correctly predicted by the model assumes a multivariate normal probability density function (PDF) for Eq.~(\ref{eq:likelihood}) with a standard deviation $\sigma$:
\begin{equation}
p({\bf d}|{\bf m},\sigma)=\frac{1}{(2\pi \sigma^2)^{N_{d}/2}}\exp\left(-\frac{1}{2 \sigma^2}\left({\bf d}-{\bf G}({\bf m})\right)^{T}\left({\bf d}-{\bf G}({\bf m})\right)\right)
\label{eq:likelihood}
\end{equation}
where $N_{d}$ is the number of measurements. The standard deviation $\sigma$ encapsulates both measurement and modelling errors and is determined here during the inversion.
 We refer to it later as the estimated 'noise' level. It has to be ultimately compared  with the typical accuracy of the picking of the diffracted waves arrival (denoted as $\sigma_d$) to quantify the modelling error. 
We assume independent the prior PDFs on the model parameters $\bf{m}$ and noise variance $\sigma$:
 $p({\bf m},\sigma)=p({\bf m})p(\sigma)$.
 The noise level can only be positive (Jeffrey's parameter). We thus invert of 
 $\beta=\ln \sigma, \beta \in \left(-\infty,\infty\right)$ and assume a uniform prior PDF for $\beta$ $p(\beta)=1$ ($p(\sigma)=1/\sigma$ - Jeffrey's prior 
 \citep{Tara05}). 
We model the prior knowledge on the model parameters as a multivariate Gaussian PDF:
\begin{equation}
    p({\bf m})=\frac{1}{\left(2\pi\right)^{N_{p}/2}|{\bf C}_{p}|^{1/2}}\exp\left(-\frac{1}{2}\left({\bf m}-{\bf m}_{p}\right)^{T}{\bf C}_{p}^{-1}\left({\bf m}-{\bf m}_{p}\right)\right)
    \label{eq:prior}
\end{equation}
$\mathbf{m}_{p}$ are prior means for the $N_p$ model parameters (see Table.~\ref{tab:Model}) where we use $A=\ln a, B=\ln b, R=\ln r$ as the fracture dimensions ($a, b, r$) must be strictly positive.
We assume that the different model parameters are a-priori un-correlated 
(${\bf C}_p$ is diagonal). 
As shown in Table.~\ref{tab:Priors},  the same priors standard deviations are taken for all the models and are chosen to be rather uninformative. The vertical position of the fracture center $x_3$ and the tilting angle of the fracture plane $\theta$ are however more constrained than the other parameters according to  post mortem analysis of the fracture plane location inside the block.

\begin{table}
\centering
\caption{Table of priors used for different models. Prior standard deviations are shown in the parentheses. \label{tab:Priors}} 
\begin{tabular}{c|c|c|c|c|c|c}
\hline
Experiment &  $A$, $B$, $R$ & $x_1$, $x_2$ (m) & $x_3$ (m)& $\psi$ (rad) & $\theta$ (rad) & $\phi$ (rad)\\
\hline
GABB-001 &  ln(0.05) (-ln(0.125)) & 0.125 (0.02) & 0.1285 (0.006) & 0 ($\pi/4$)  & 0 ($\pi/60$) & 0 ($\pi/2$)  \\
MARB-005 &  ln(0.05) (-ln(0.125)) & 0.125 (0.02) & 0.131 (0.013) & 0 ($\pi/4$)  & 0 ($\pi/40$) & 0 ($\pi/2$) \\
\hline
\end{tabular}
\end{table}

Using Bayes theorem and considering the probability of the data being observed as a normalizing constant, introducing $\mathbf{z}=(\mathbf{m},\,\beta)$ we write the normalized posterior PDF as $ \Pi(\mathbf{z}|\mathbf{d})=p(\mathbf{d}|\mathbf{z})p(\mathbf{m})p(\beta)$. Several techniques can be used to quantify such a posterior PDF - e.g. from global Markov-Chain Monte Carlo (MCMC) sampling, to local quasi-Newton searches. Our aim here is to seek the most probable solution (PDF mode) and estimate the posterior uncertainties around this solution.
This is equivalent to finding the minimum of $-\text{ln }\Pi(\mathbf{z}|\mathbf{d})$. The posterior uncertainties can be grasped either via direct Monte Carlo sampling or in a cheaper way by approximating the posterior PDF as a multivariate Gaussian near its mode:
\begin{equation}
   \Pi(\mathbf{m},\sigma_m|\mathbf{d})=\Pi(\mathbf{z}|\mathbf{d})\approx\Pi(\mathbf{\tilde{z}}|\mathbf{d})\text{exp}(-\frac{1}{2}(\mathbf{z}-\mathbf{\tilde{z}})^{T}\tilde{\mathbf{C}}^{-1}(\mathbf{z}-\mathbf{\tilde{z}}))
   \label{eq:Posterior}
\end{equation}
where $\mathbf{\tilde{z}}$ represents the most probable model parameters and modelling noise and $\mathbf{\tilde{C}}$ the posterior covariance matrix at $\mathbf{\tilde{z}}$. 

We have applied different algorithms to estimate $\mathbf{\tilde{z}}$ here. 
For robustness, although a simple local quasi-Newton scheme is sufficient for most cases, we present results obtained using a global minimization algorithm (direct differential evolution  \citep{StPr97}). The posterior covariance matrix is estimated from the Hessian of $-\text{ln }\Pi(\mathbf{z}|\mathbf{d})$ at $\mathbf{\tilde{z}}$. We have notably compared  the posterior mode and uncertainties with the results obtained from  MCMC sampling of the posterior PDF $\Pi(\mathbf{m},\sigma_m|\mathbf{d})$. The results are similar.

\subsection{Bayes factor}

The data are inverted with the different geometrical models listed in Table \ref{tab:Model}.
The selection of the most suitable model is drawn not only from the quality of fit, but must also account for model complexity.
We use Bayes factor to rank between two possible models assuming equi-probable models a-priori.
The Bayes factor between model $i$ and $j$ is defined as the 
ratio of the marginal probability of the data for the given model \citep{raftery1995hypothesis}:
\begin{equation}
B_{ij}=\frac{p({\bf d|{\bf \mathcal{M}_{i}}})}{p({\bf d|{\bf \mathcal{M}_{j}}})}
\label{eq:Bayesfactor}
\end{equation}
where the marginal probability of the data $p(\mathbf{d|\mathcal{M}})$ is obtained by integrating the posterior PDF for the given model over the complete model parameters space: 
\begin{equation}
p(\mathbf{d|}\mathcal{M})=\int_{\mathbf{m}_k} \int_{\sigma_k} \Pi(\mathbf{d}|\mathbf{m}_{k},\sigma_{k})\text{d}\mathbf{m}_{k}\text{d}\sigma_{k}
\label{eq:modelpr}
\end{equation}
To obtain such a probability by cheaper means than Monte Carlo sampling, 
we approximate the posterior PDF around the most probable value as a multivariate Gaussian (see Eq.(\ref{eq:Posterior})).
We thus estimate the marginal probability of the model (\ref{eq:modelpr}) as
\begin{equation}
    p(\mathbf{d}|\mathcal{M})\approx\Pi(\mathbf{\tilde{z}}|\mathbf{d})(2\pi)^{(N_{p}+1)/2}|\tilde{\mathbf{C}}|^{1/2}
    \label{eq:PosteriorModel}
\end{equation} 
As noted in \cite{raftery1995hypothesis}, for a Bayes factor $B_{ij}>10$, the data clearly favours the model $\mathcal{M}_i$ over the model $\mathcal{M}_j$ (respectively $\mathcal{M}_j$ over $\mathcal{M}_i$ for  $B_{ij}< 0.1$). For a Bayes factor between $0.2$ and $5$, the models are equivalent.

\section{Results and discussions}

\begin{figure}
\centering 
\includegraphics[height=0.30\linewidth]{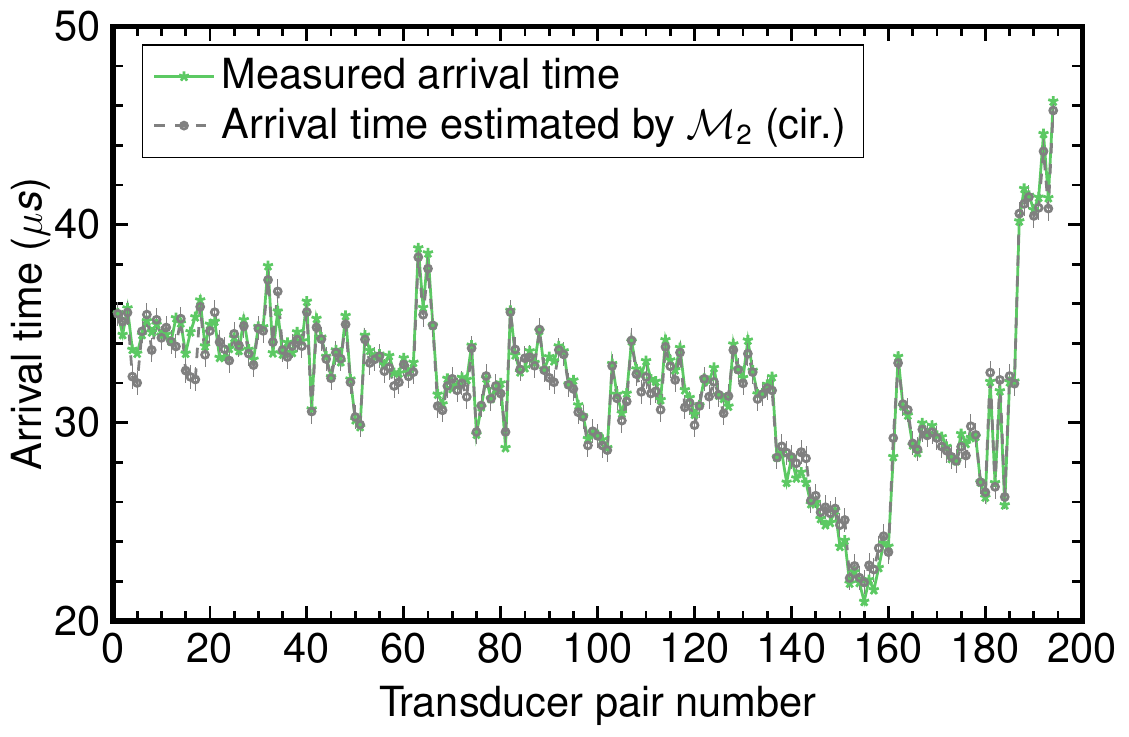}
\includegraphics[height=0.30\linewidth]{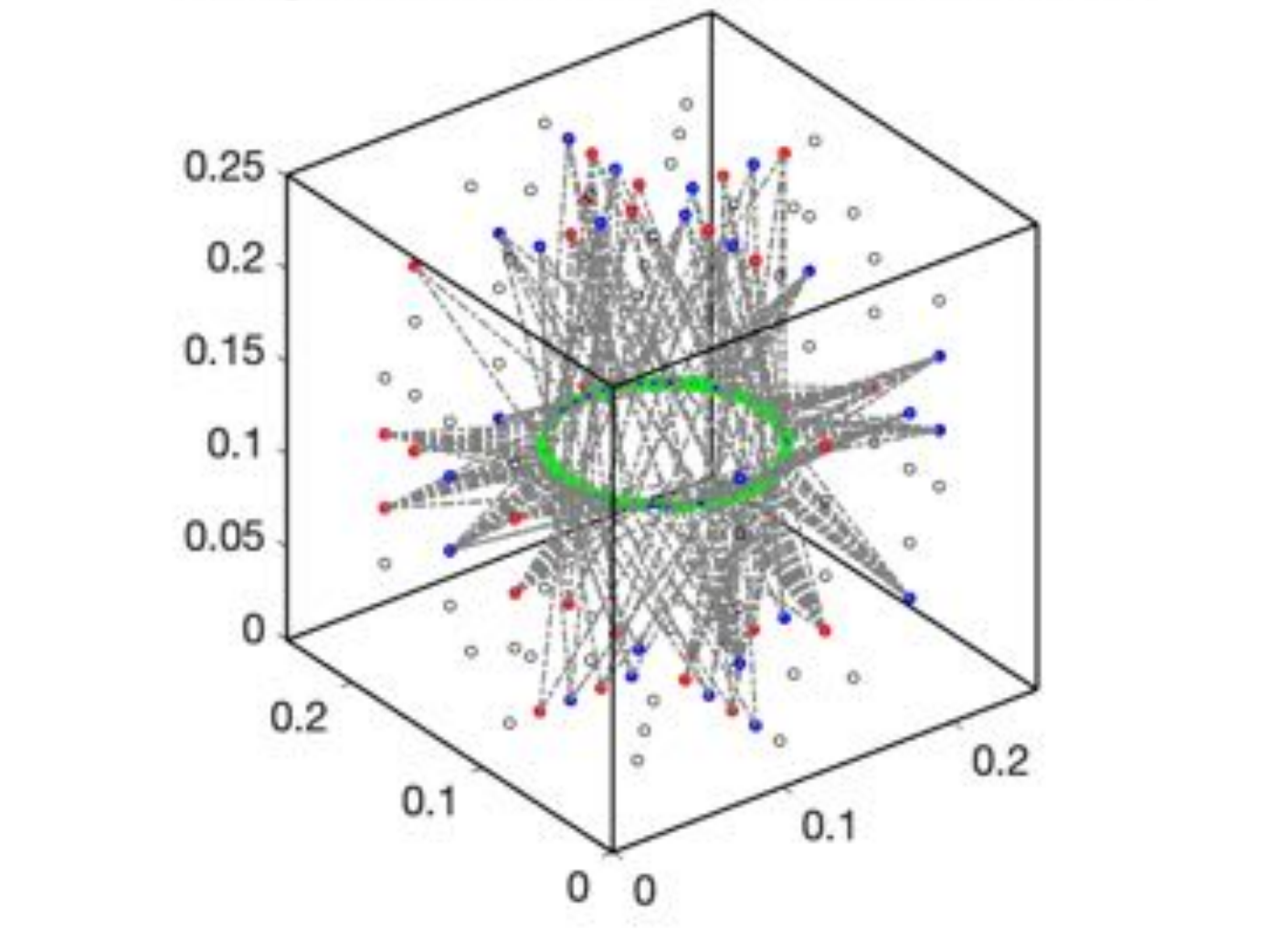}
\caption{Comparison of the measured arrival and predicted arrival time for model $\mathcal{M}_2$ -  sequence 50 in GABB-001 experiment and illustration of diffractors (in green) along the fracture front with its corresponding travel path (in gray) for acoustic diffraction. The red and blue points represent respectively the sources and receivers.}
\label{fig:Illstration}
\end{figure}

The number of source-receiver pairs with picked diffracted arrivals vary between acquisition sequences, type of diffracted waves (fracture tip PdP, fluid front PdP and PdS) and experiments. Except for early sequences, the data pair is always larger than 100 (see Fig.~\ref{fig:PickedPairsNumber}).
For one chosen model and a given sequence, we perform the inversion using all the picked diffracted arrivals.
Fig.~\ref{fig:Illstration} displays an example of the  model predictions and data (left panel) as well as the corresponding fracture front and diffracted waves ray path for that sequence (right panel).
We repeat the inversion procedure for each sequence and obtain the evolution of the fracture, fluid front together with their posterior covariances. 
We do this for all four forward geometric models in Table~\ref{tab:Model} and rank them using the estimated Bayes factors. We also compare the 
 noise level $\sigma$ estimated with the picking accuracy of the diffracted wave arrivals $\sigma_d$. A large difference between $\sigma$ and $\sigma_d$ (about one order of magnitude) indicates that the chosen model is clearly not capable of properly reproducing 
 the data.

\subsection{Toughness dominated experiment GABB-001}
\begin{figure}
\centering
\begin{tabular}{cc}
\includegraphics[height=0.33\linewidth]{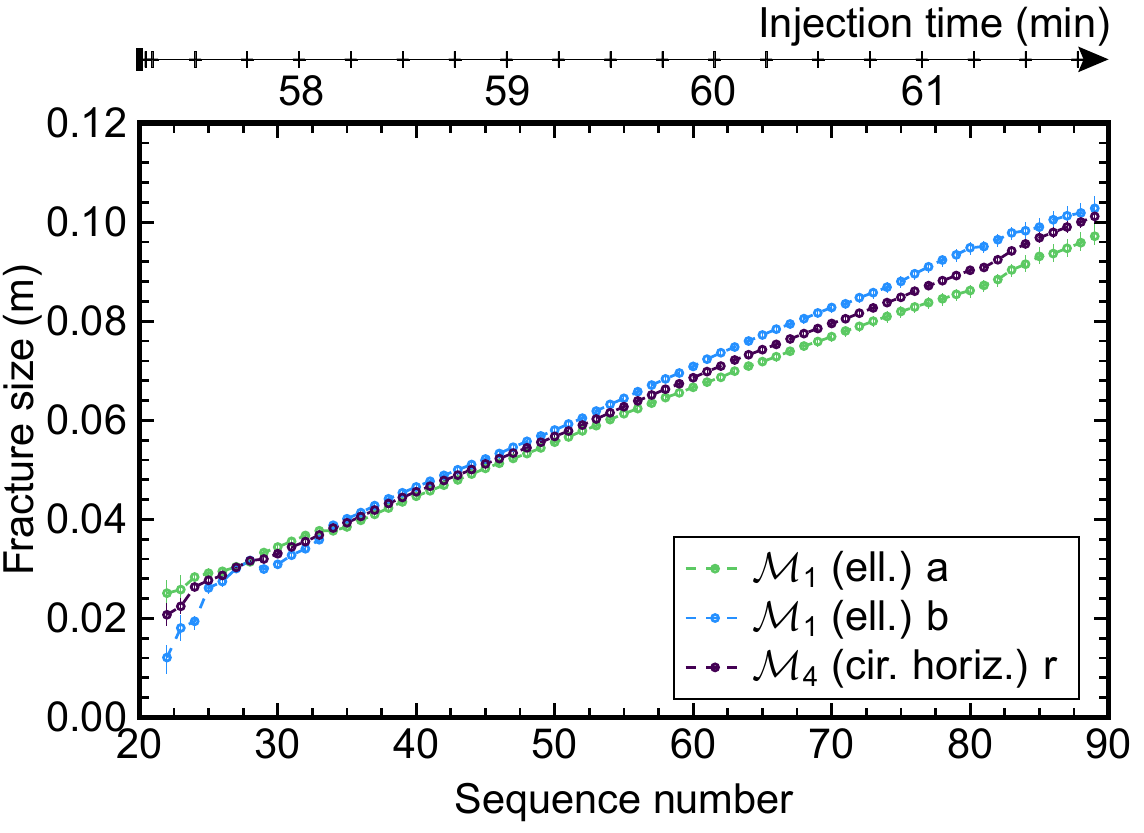}&
\includegraphics[height=0.33\linewidth]{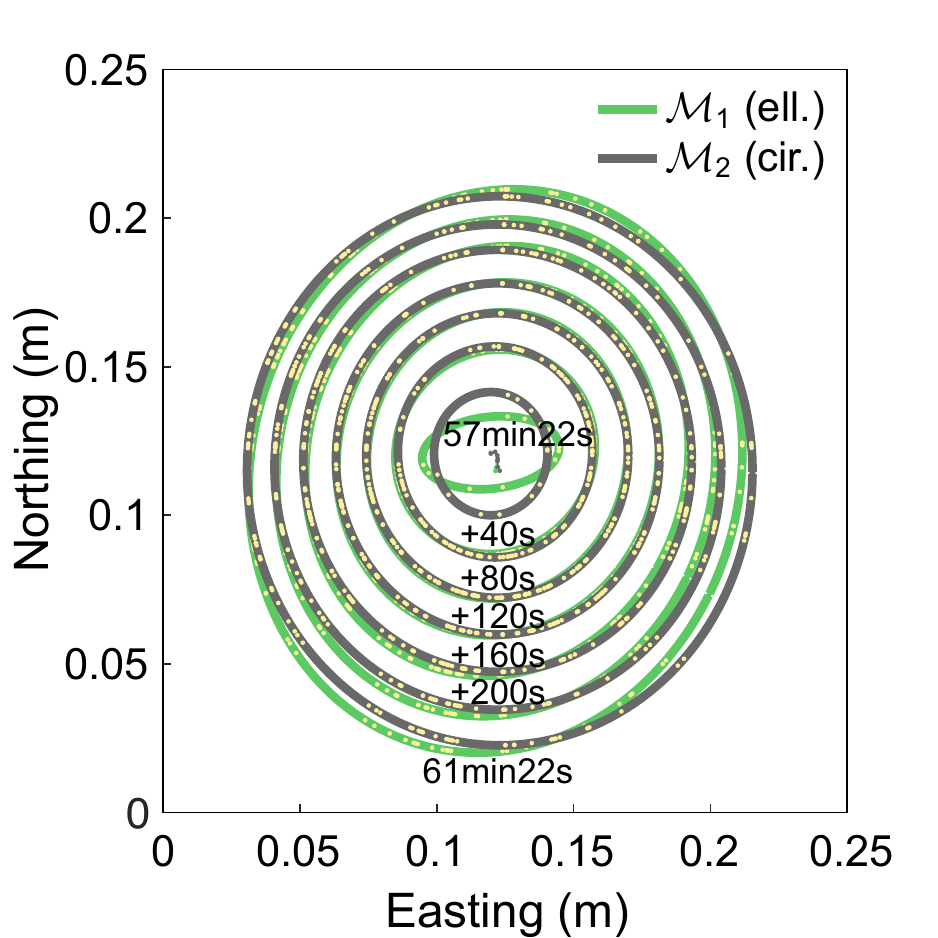}
\\
(a)& (b)\\
\includegraphics[height=0.33\linewidth]{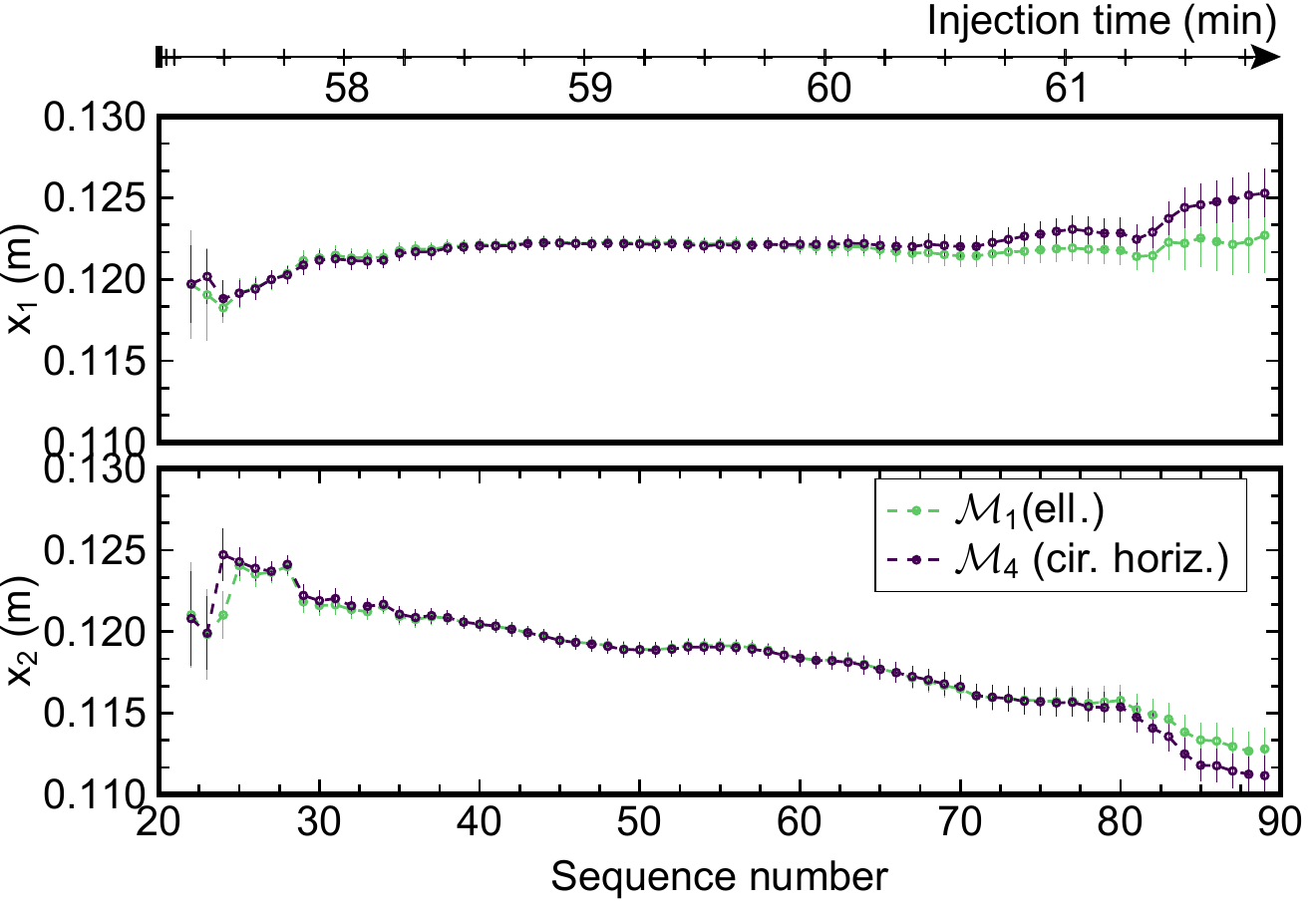} &
\includegraphics[height=0.33\linewidth]{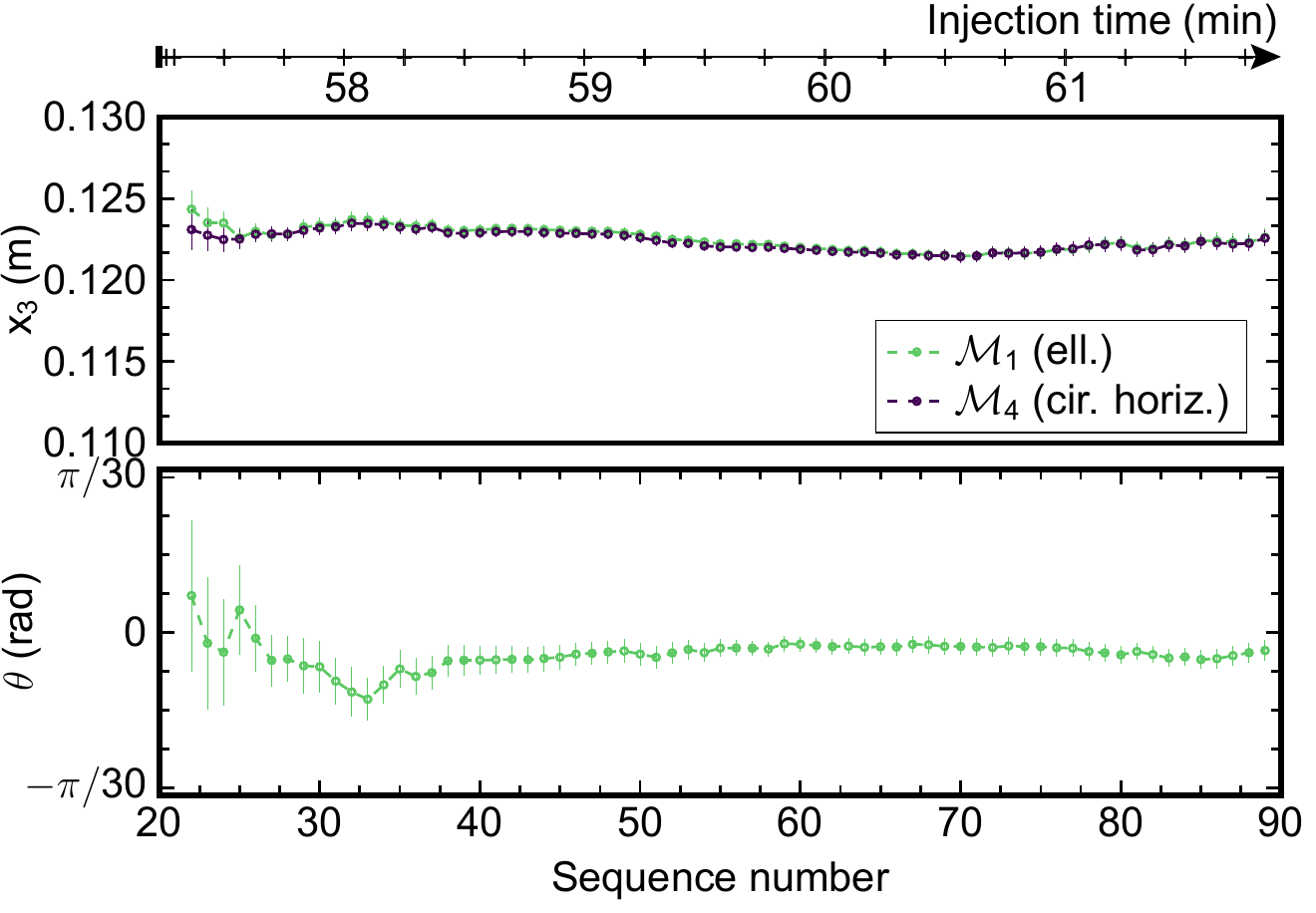}\\
(c) & (d)\\
\end{tabular}
\caption{GABB-001 experiment: evolution of the fracture size (a), offset of the fracture center (c) and tilt of the fracture plane (d). The figure (b) displays the footprint of the fracture from a top view (from sequence 22 to sequence 82) shown every 10 sequences. The yellow dots in (b) indicate the diffractors at the fracture front for the different source receiver pairs picked.}
\label{fig:G01FractureSize}
\end{figure}

\begin{figure}
\centering
\begin{tabular}{cc}
\includegraphics[height=0.33\linewidth]{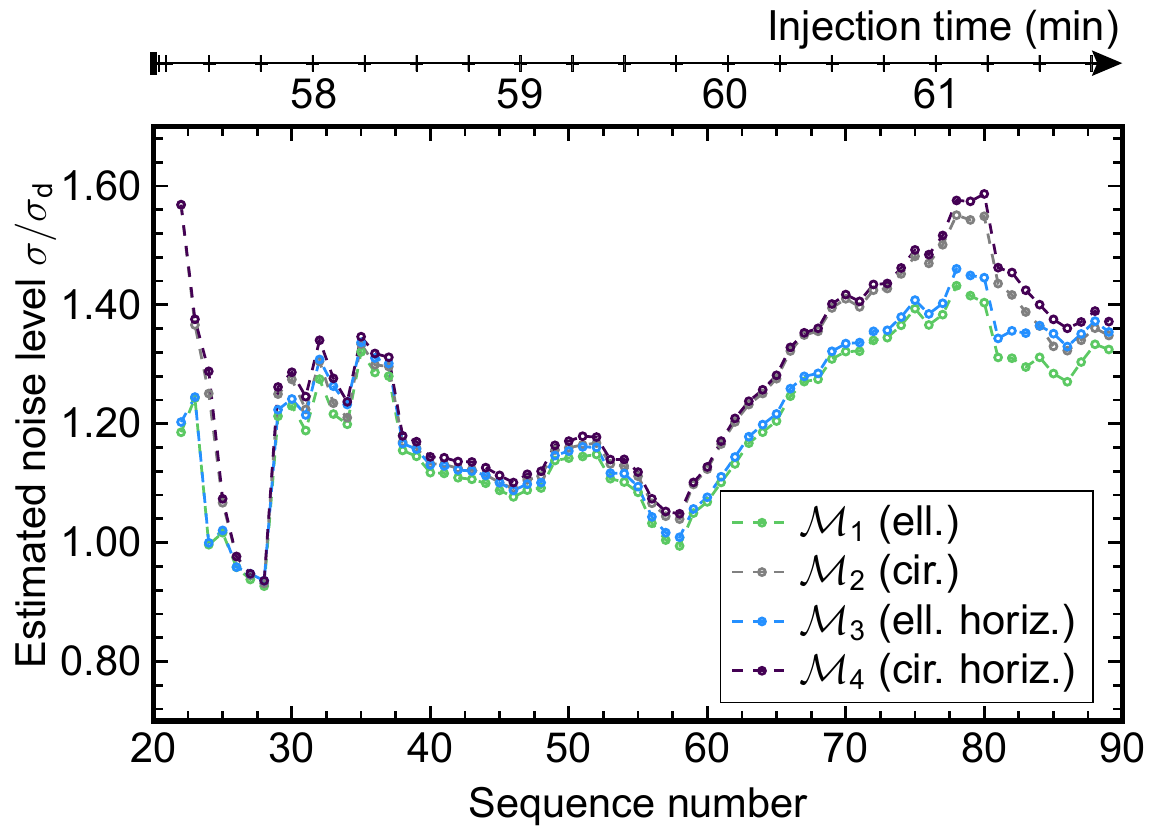} &
\includegraphics[height=0.33\linewidth]{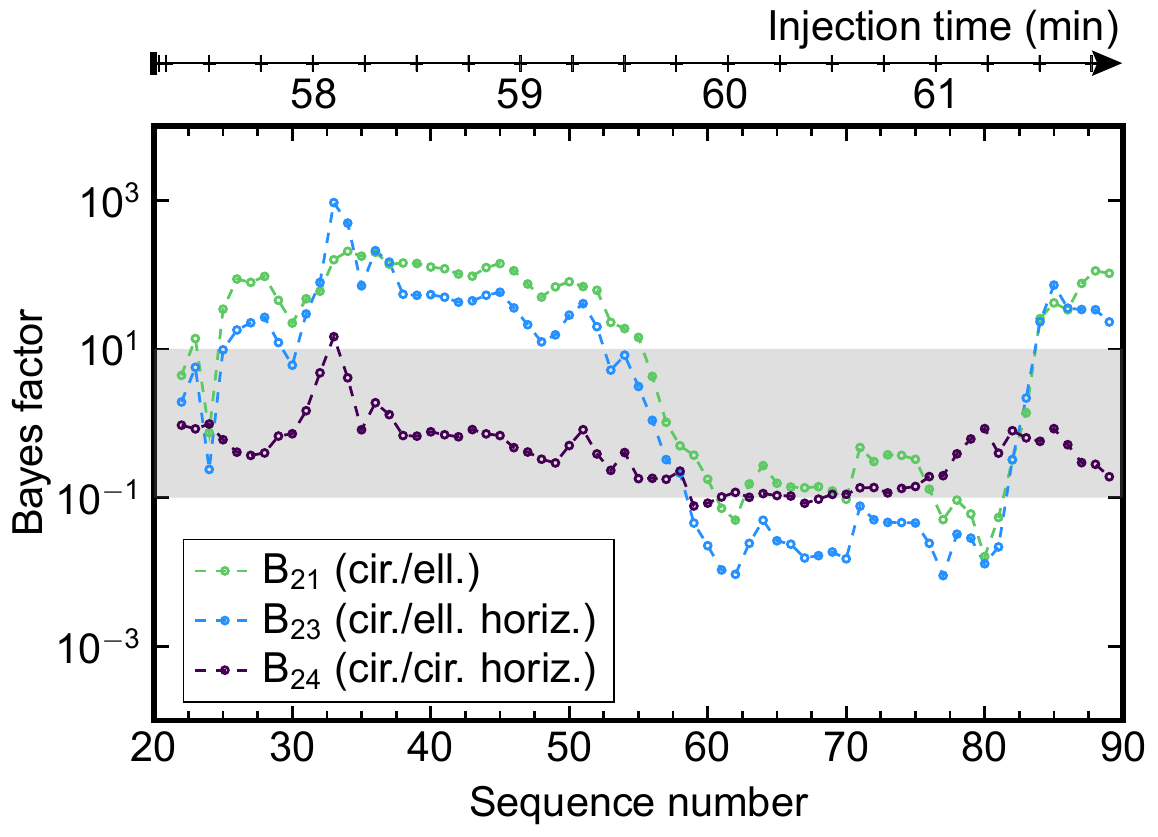}
\end{tabular}
\caption{GABB-001 experiment: evolution of estimated noise level (left, the estimated picking error $\sigma_d=0.5$~$\mu$s in GABB-001) and Bayes factor (right). The gray region characterises $0.1<B_{ij}<10$, where $\mathcal{M}_i$ and $\mathcal{M}_j$ can not be decisively ranked.}
\label{fig:G01ModelComparison}
\end{figure}

The GABB-001 experiment presents a steady fracture growth throughout the block as illustrated in Fig.~\ref{fig:G01FractureSize}. This is in line with the steady increasing entering flux shown in Fig.~\ref{fig:G01NonAcoustic} as previously discussed.
No fluid lag was observed from the acoustic diffraction data during the fracture growth. 
Larger posterior uncertainties and estimated model noises (Fig.~\ref{fig:G01ModelComparison}) are found for early sequences due to the limited number of picked arrivals pairs (see Fig.~\ref{fig:PickedPairsNumber}). 

\begin{figure}
    \centering
    \includegraphics[height=0.33\linewidth]{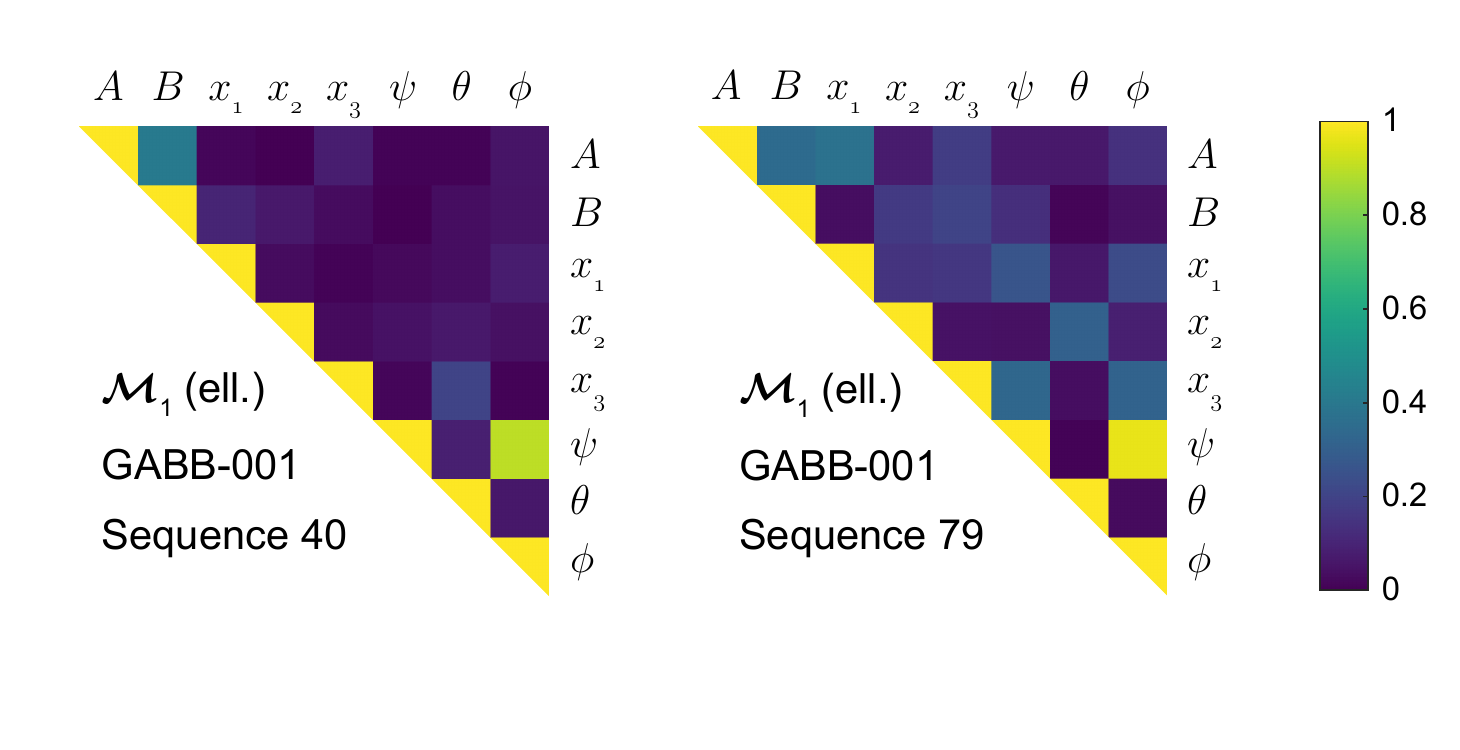}
    \caption{GABB-001 experiment: posterior correlation (absolute value) between $\mathcal{M}_1$ (ellipse) model parameters corresponding respectively to the early time and the mid-to-late stage of fracture growth.}
    \label{fig:Correlation}
\end{figure}

All four geometrical models provide a good fit to the data with an estimated noise level of the same order of magnitude as the estimated manual picking accuracy $\sigma_d=0.5$~$\mu$s as shown in Fig.~\ref{fig:G01ModelComparison}. 
From the Bayes factor,
the fracture shape appears to be better described by a radial geometry than an elliptical one, particularly during the first half stage of the propagation as $B_{21}, B_{23}>10$ (see Fig.~\ref{fig:G01ModelComparison}). 
The strong posterior correlation between the two semi-lengths for the elliptical model $\mathcal{M}_1$ (Fig.~\ref{fig:Correlation}) and
an aspect ratio around 1   (Fig.~\ref{fig:AspectRatioEvolution})
confirms the preference for the circular geometry.
The latter stage of the fracture propagation 
presents a slightly larger estimated noise level (Fig.~\ref{fig:G01ModelComparison}) even though the number of 
picked arrival remains large (see Fig.~\ref{fig:PickedPairsNumber}). 
This hints that the chosen models start to become inadequate at later times. 
This is most likely due 
to the non-uniformity of the stress field near the bock edges which results in a deviation of the fracture geometry from a circular / elliptical shape. Over that period, a  drop of $B_{21}, B_{23}$ in Fig.~\ref{fig:G01ModelComparison} can be observed. The inversion then slightly favours the elliptical fracture models although their estimated noise level increase similarly than the radial models (Fig.~\ref{fig:G01ModelComparison}).

The fracture plane remains approximately horizontal during the whole fracture growth with a dip fluctuating around zero. This horizontal geometry is further confirmed by the Bayesian analysis. Given that $B_{24}$ remains in the range $0.1-10$ most of the time, the model ($\mathcal{M}_2$) allowing for a possible tilt of the fracture plane is nearly equivalent to the strictly horizontal one ($\mathcal{M}_4$) for the same radial fracture geometry. The fracture center deviates little in the vertical direction through the entire propagation (Fig.~\ref{fig:G01FractureSize}). As a result, strong posterior correlations among the different Euler angles as well as with the fracture center coordinates are observed for the elliptical model (Fig.~\ref{fig:Correlation}).

\subsection{Lag-viscosity dominated experiment MARB-005}

\begin{figure}
\centering 
\begin{tabular}{cc}
\includegraphics[height=0.33\linewidth]{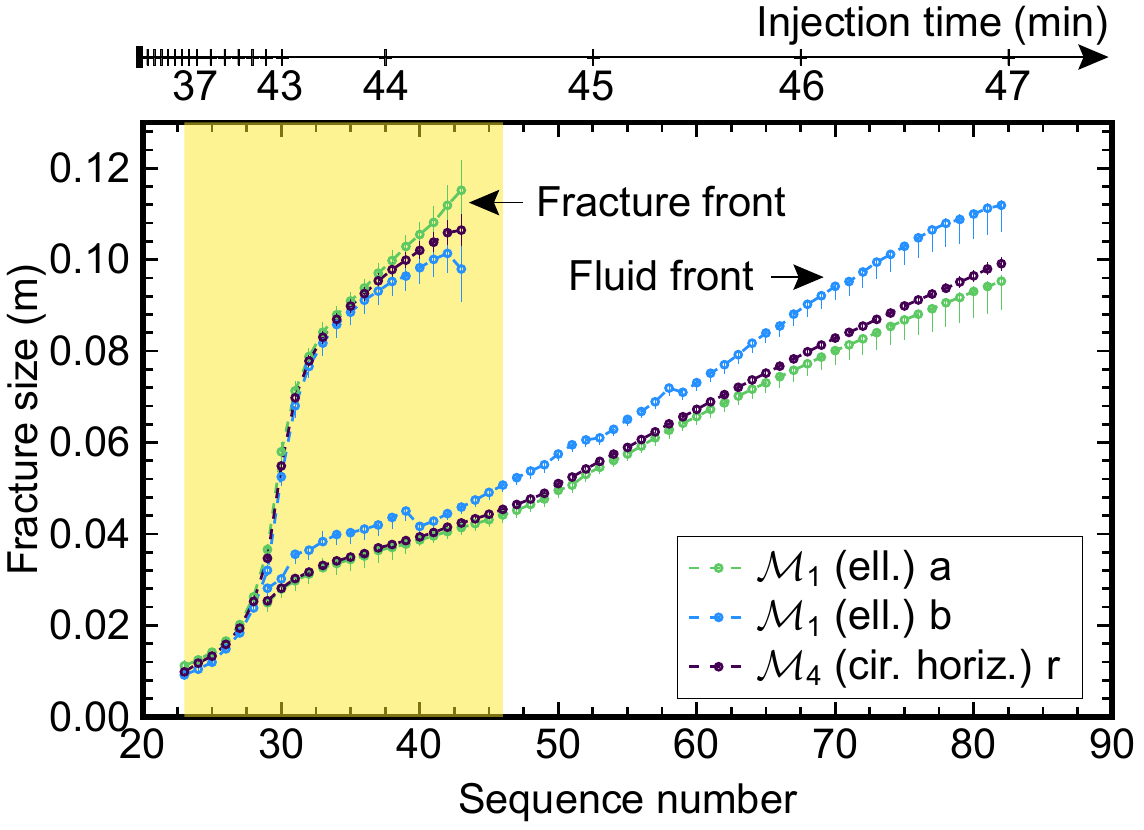}&
\includegraphics[height=0.33\linewidth]{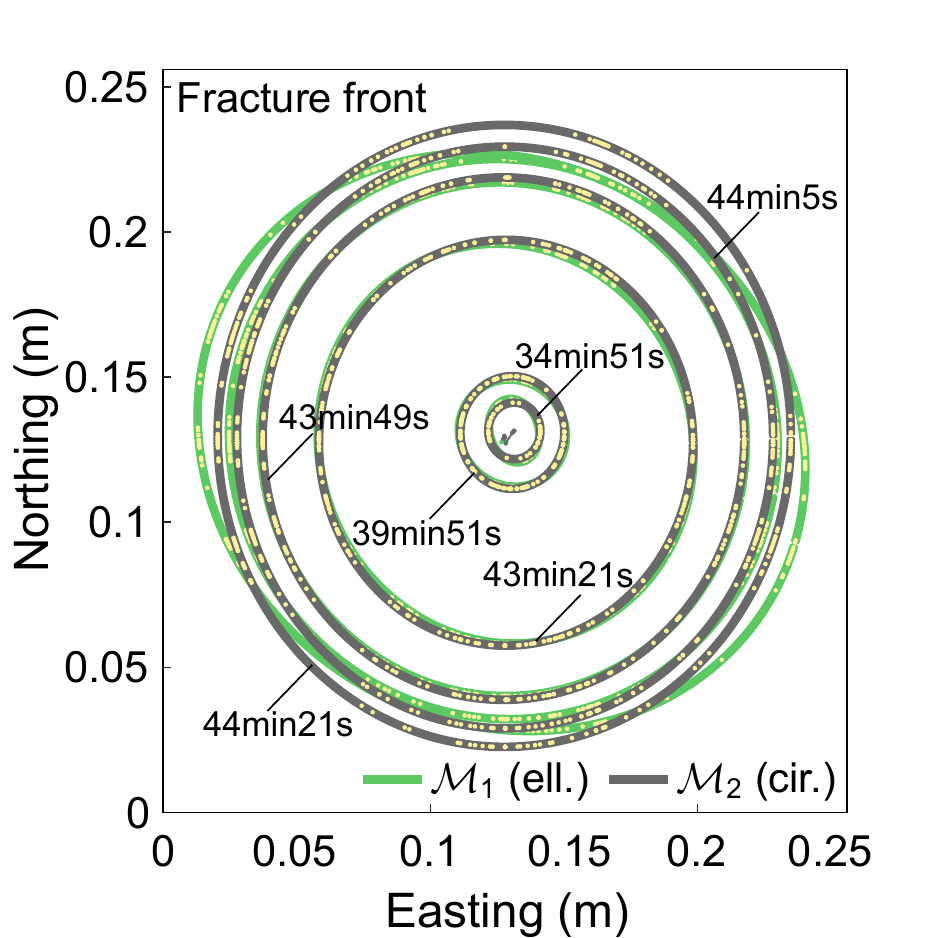}\\
(a) & (b)\\
\includegraphics[height=0.33\linewidth]{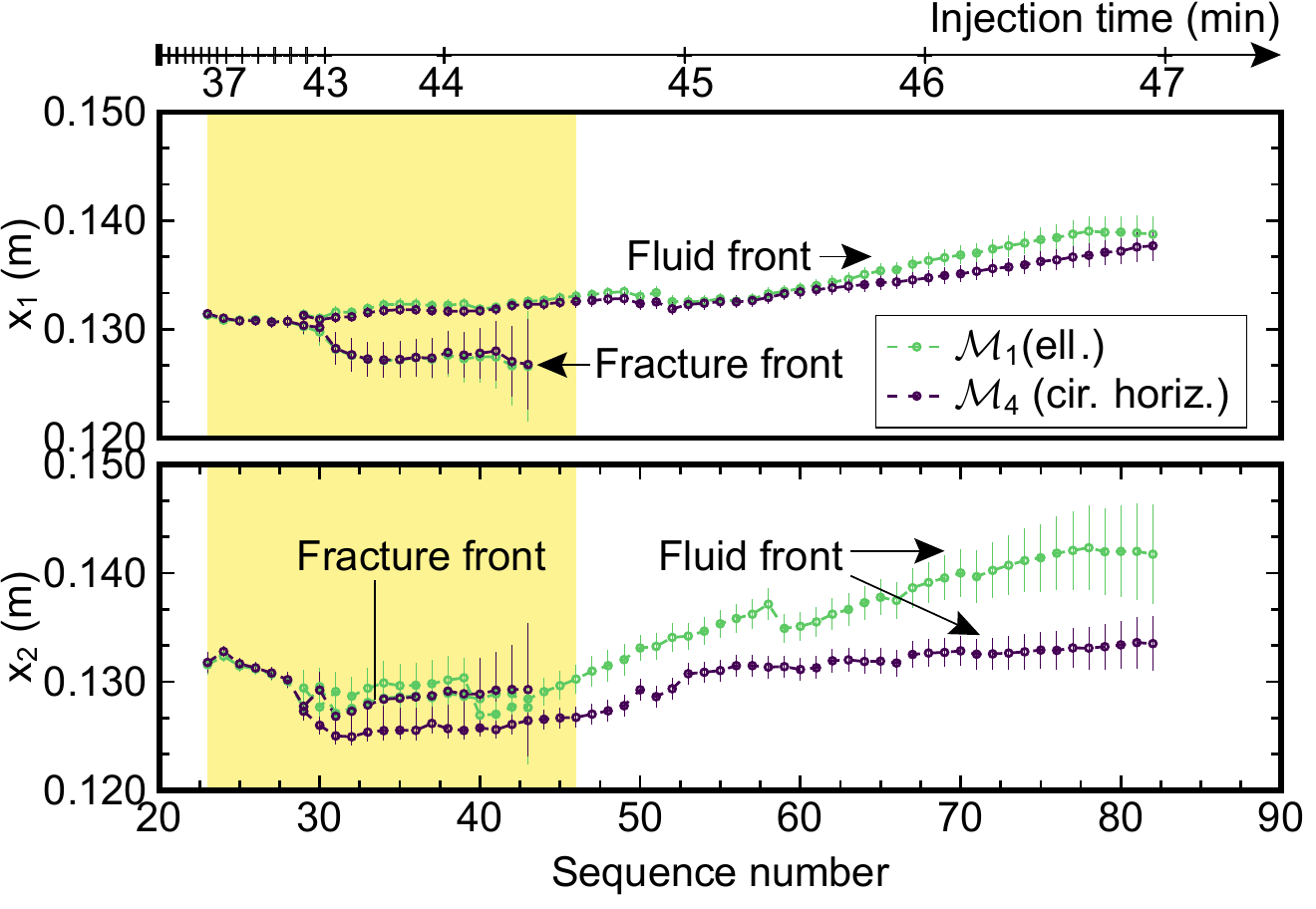}&
\includegraphics[height=0.33\linewidth]{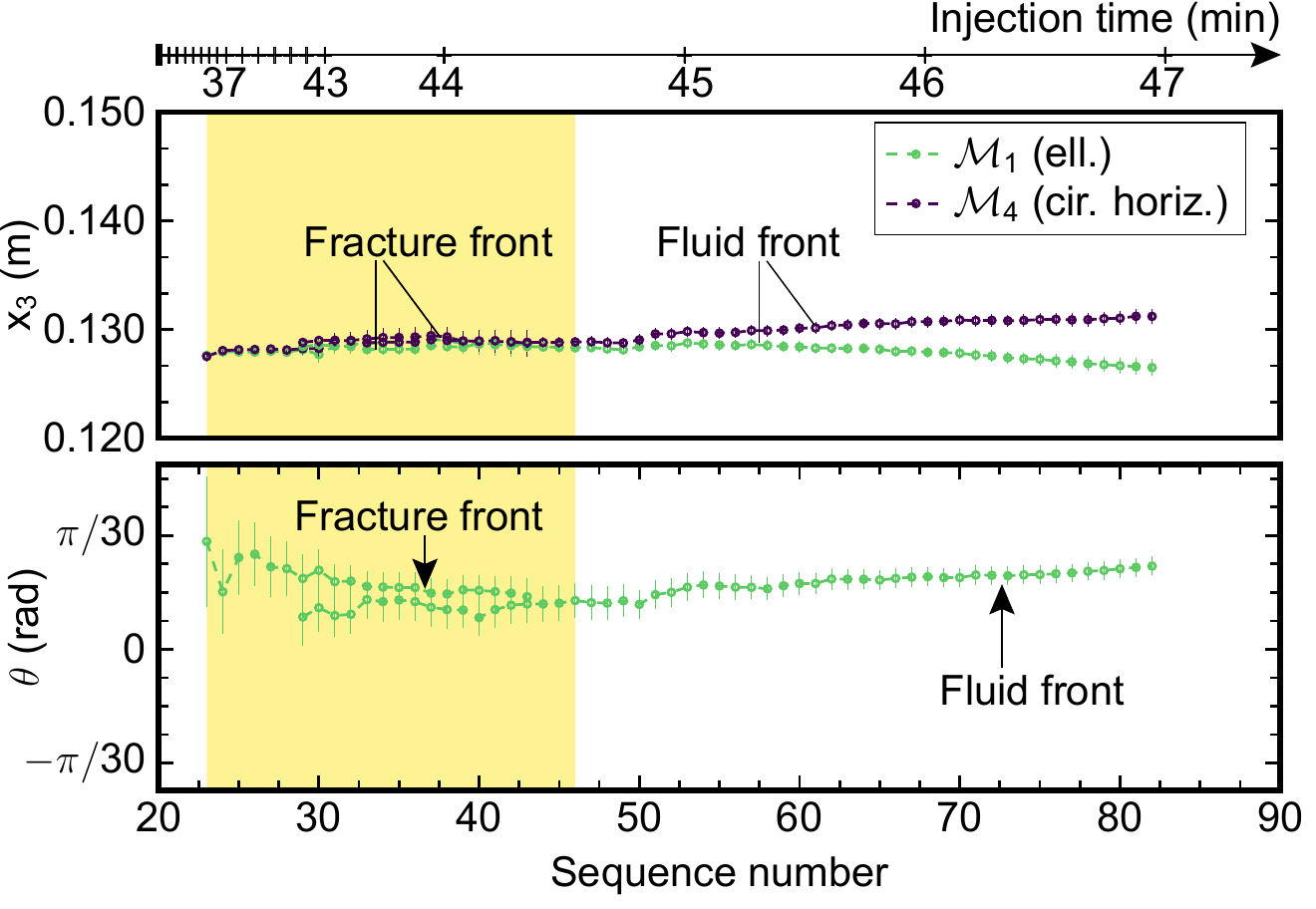}\\
(c) & (d)\\
\end{tabular}
\caption{MARB-005 experiment: evolution of fracture size (a), fracture center offset (c) and the tilt of the fracture plane (d). The yellow coloured time interval represents the propagation of the fracture through the specimen. The figure (b) displays the footprint of the fracture front and the fracture center from the top view  (from sequence 23 to sequence 43, shown every 4 sequences). The yellow dots in (b) indicate the diffraction points at the fracture front for the different source receiver pairs picked.}
\label{fig:M05FractureSize}
\end{figure}

\begin{figure}
\centering 
\includegraphics[height=0.33\linewidth]{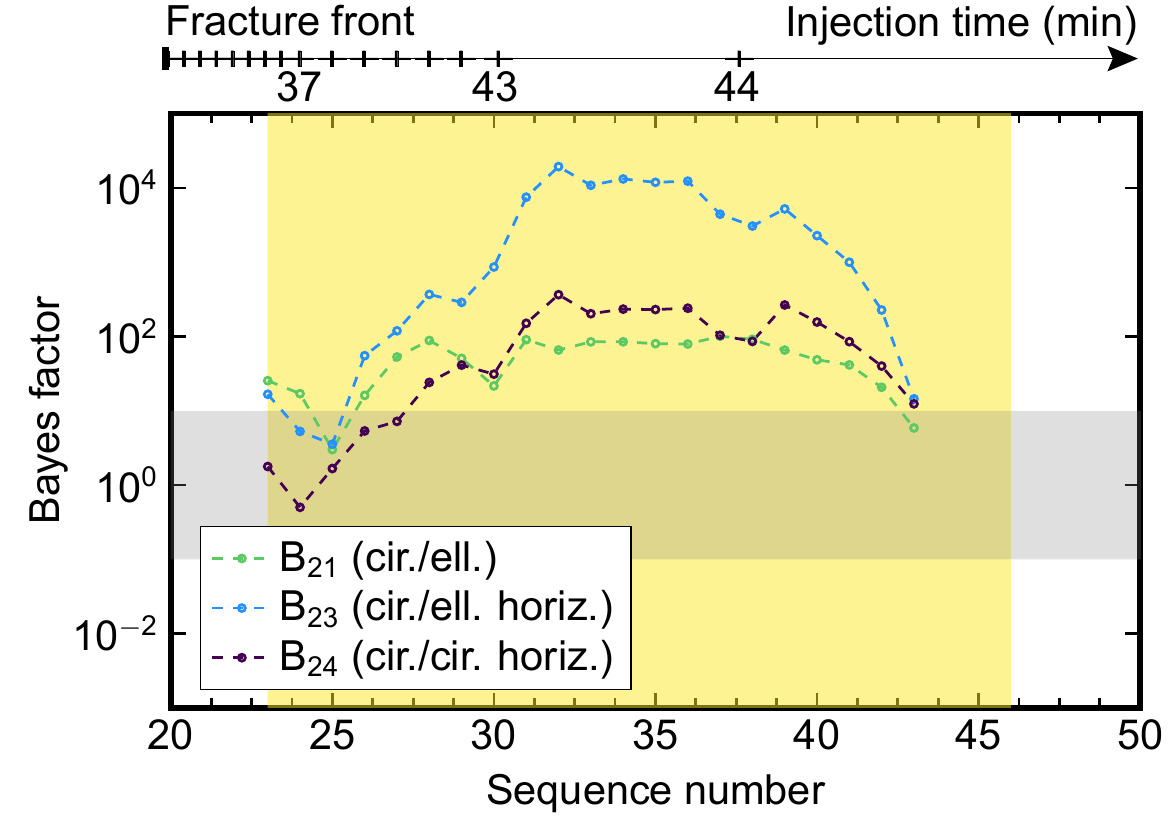}
\includegraphics[height=0.33\linewidth]{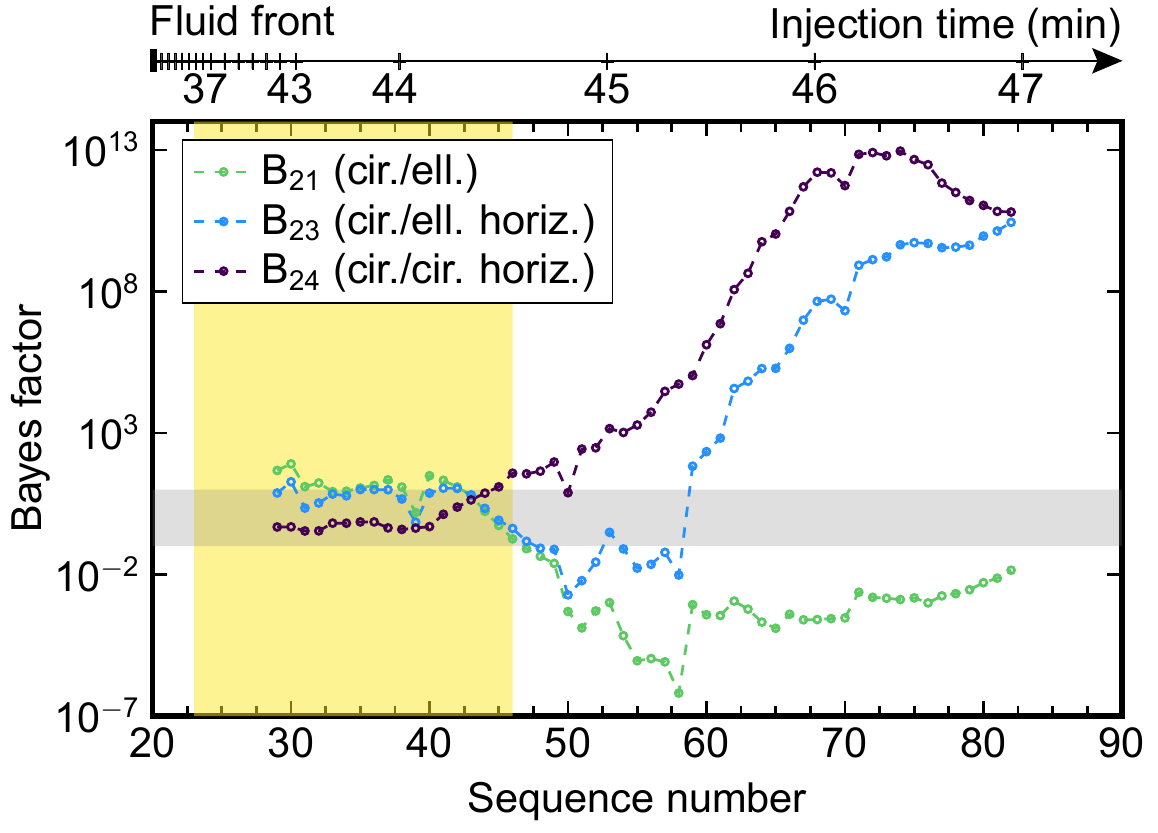}
\caption{MARB-005 experiment: evolution of Bayes factor for the fracture front (left) and the fluid front (right). The yellow coloured time interval represents approximately the propagation of the fracture from initiation to the edges of the block. The gray region characterises $0.1<B_{ij}<10$, where $\mathcal{M}_i$ and $\mathcal{M}_j$ are considered equivalent in fracture geometry description.}
\label{fig:MBayes}
\end{figure}

\begin{figure}
\centering 
\includegraphics[height=0.33\linewidth]{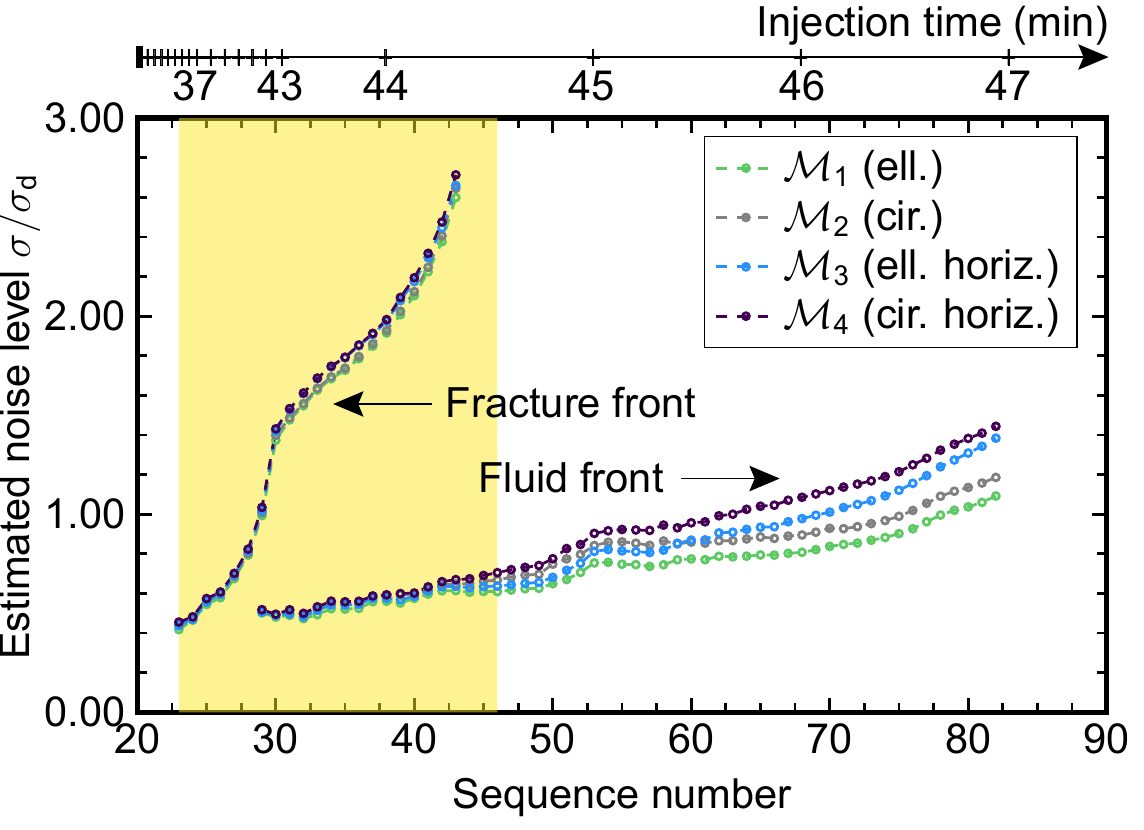}
\includegraphics[height=0.33\linewidth]{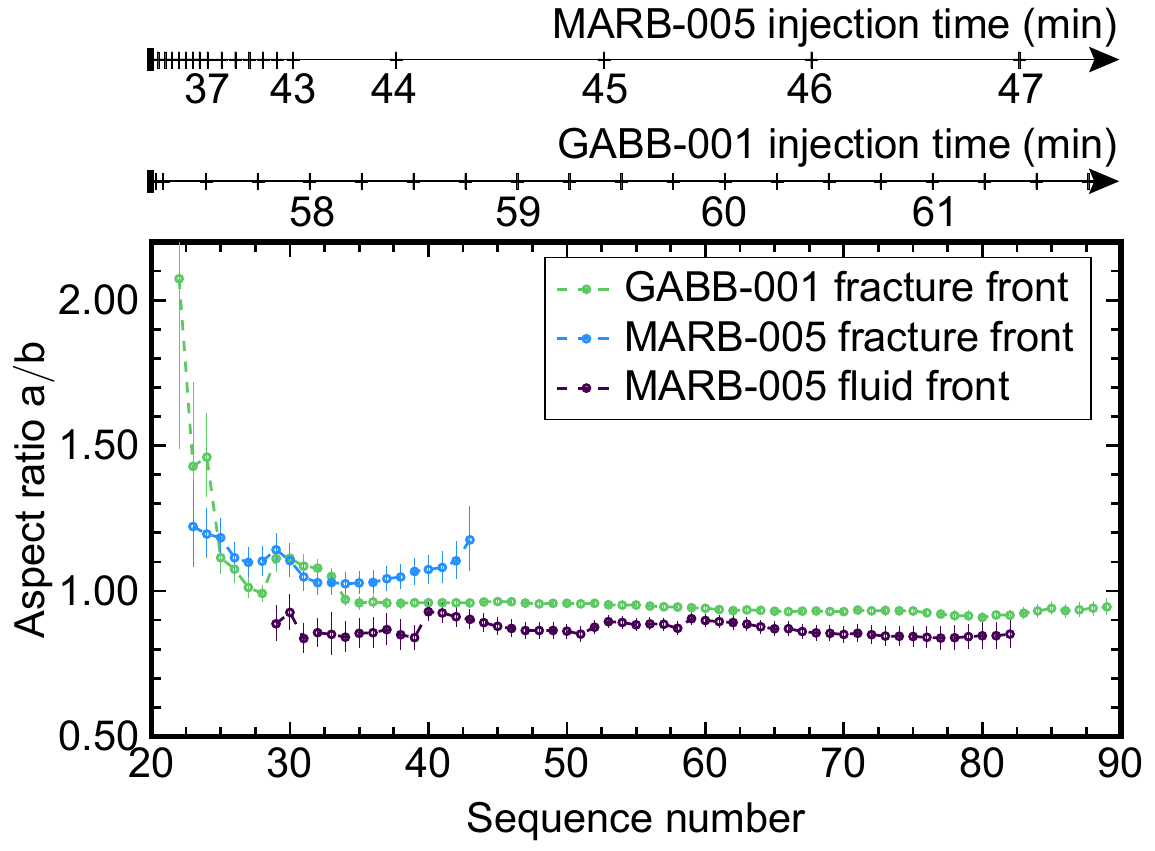}
\caption{Evolution of the modelling noise in MARB-005 (left, the estimated picking error $\sigma_d=1$~$\mu$s in MARB-005). The yellow coloured time interval  represents approximately the propagation of the fracture from initiation to the edges of the block. Evolution of aspect ratio (right) in GABB-001 and MARB-005 experiments by assuming an elliptical fracture geometry ($\mathcal{M}_1$).}
\label{fig:AspectRatioEvolution}
\end{figure}

Fig.~\ref{fig:M05FractureSize} shows a fast growth of the fracture front followed by a gradual evolution of the fluid front. 
Due to the strong viscous effect, we observe a continuous increase of the fluid pressure even after fracture initiation. The entering flow rate remains very small up to around 8 minutes after initiation (Fig.~\ref{fig:M05NonAcoustic}). It then increases significantly when the fracture reaches the edge of the block.

During most of fracture growth, the fracture front geometry is better described by the circular model $\mathcal{M}_2$ as $B_{21}, B_{23}, B_{24}>10$ (see Fig.~\ref{fig:MBayes}). 
The radial geometry is favoured for the fluid front shortly after  fracture initiation. The  tilted models and particularly the elliptical tilted model (for the fluid front) become more probable after 44 to 45 minutes of injection which corresponds to the time when the fracture front reaches the end of the block (see Fig.~\ref{fig:MBayes}).

The fracture plane does not remain horizontal during the fracture growth as illustrated in Fig.~\ref{fig:M05FractureSize}. This is in line with the evolution of the Bayes factors with $B_{23}, B_{24}>10$ for most sequences (Fig.~\ref{fig:MBayes}). The tilt of the fracture plane has been confirmed by postmortem observations with an averaged tilt of around $\pi/50$ ($3.6^\circ$). The fluid front however evolves differently from the fracture front, characterised by a different size and center (Fig.~\ref{fig:M05FractureSize}). At  early time,  no decisive discrimination can be made between the tilted plane models and horizontal ones. After 44 to 45 minutes of injection when the fracture front reaches the edge, the fluid flows much more freely between the fractured surfaces and the fluid front geometry tends to favour the tilted models given the increase of $B_{23}$ and $B_{24}$. 

The quality of the fit to data is acceptable. The estimated modelling noise is of the same order of magnitude as the picking error $\sigma_d=1$~$\mu$s in Fig.~\ref{fig:AspectRatioEvolution}. The fluid front presents a lower noise level  even though  less picked arrivals were used in the inversion  (Fig.~\ref{fig:PickedPairsNumber}). This explains a better fit of the predicted arrivals to the picked ones for the fluid front. In the latter half of the fluid front propagation, we observe an important difference between models and their noise level $\mathcal{M}_1<\mathcal{M}_2<\mathcal{M}_3<\mathcal{M}_4$. This is consistent with the estimation of Bayes factors which indicates also a preference for $\mathcal{M}_1$ over $\mathcal{M}_2$, $\mathcal{M}_3$, $\mathcal{M}_4$  (Fig.~\ref{fig:MBayes}) during the same period.

\subsection{Comparison with acoustic transmission data}
We now compare our estimation of the fracture (and fluid for MARB-005)  front position using transmitted waves.
Transmitted waves exhibit an increase in arrival time and attenuation when passing through a fracture. The attenuation of transmitted waves appears when the fracture front crosses the line between facing source receiver transducers located on opposite platens. It can thus be compared with the estimation from diffracted waves. 

We evaluate a transmitted "energy" by computing the signal strength $E_i^{1/2}$ of a given wave arrival (P or S) for the acquisition sequence $i$
as 
\begin{equation}
    E_i^{1/2}=\sqrt{\sum_{j=j_{min}}^{j_{max}} u_{i}^2(t_j)}
    \label{eq:transmissionenergy}
\end{equation}
where  $u_i(t)$ is a low-pass  filtered (at 2MHz) waveform which is cropped by a tapered Hamming window centered on the interest arrival with a size of $(t_{j_{max}}-t_{j_{min}})=14 \mu$s. We then choose a reference signal obtained before fracture initiation and define the attenuation ratio $(E_i/E_{ref})^{1/2}$.

An alternative is to follow the procedure presented in
\cite{groenenboom1998monitoring}
to estimate the fracture width from 
transmitted waves.
 In the frequency domain, the transmitted compressional signal $\hat{u}_{fracture}$ can be compared to the prediction obtained by
 the product of a transmission coefficient $T(\zeta,\,w)$ for 
  a three layers model (rock-fluid-rock) 
 and a reference signal $\hat{u}_{base}$ recorded before fracture initiation. The fluid layer thickness (fracture width) $w$ is estimated by minimizing the misfit for frequencies around the central frequency of the source signal ($\zeta_{min}<\zeta<\zeta_{max}$).
In addition to frequency, the transmission coefficient $T(\zeta,\,w)$ depends on solid and fluid properties (acoustic impedance) and the layer of fluid (the fracture thickness). The method compares well with optical measurements for the case of waves arriving at a 90 degree incident angle to the fracture \citep{kovalyshen2014comparison}.
We apply this method to the toughness-dominated GABB-001 experiment as the fracture is flat (thus ensuring a 90 degree incident angle) and do not exhibit any fluid lag (we take $\rho=1260  \text{kg}/\text{m}^3$ and $V_{p }=1960 \text{m/s}$ for glycerol). We set the lower and upper
 frequency bounds as $\zeta_{min}=500$~kHz and $\zeta_{max}=1100$~kHz given the central frequency of $750$~kHz. 
 We do not use such a method for the marble experiment (MARB-005) which exhibits a very large fluid lag.

\subsubsection{Lag-viscosity dominated experiment MARB-005}

\begin{figure}
    \centering
    \includegraphics[width=0.48\linewidth]{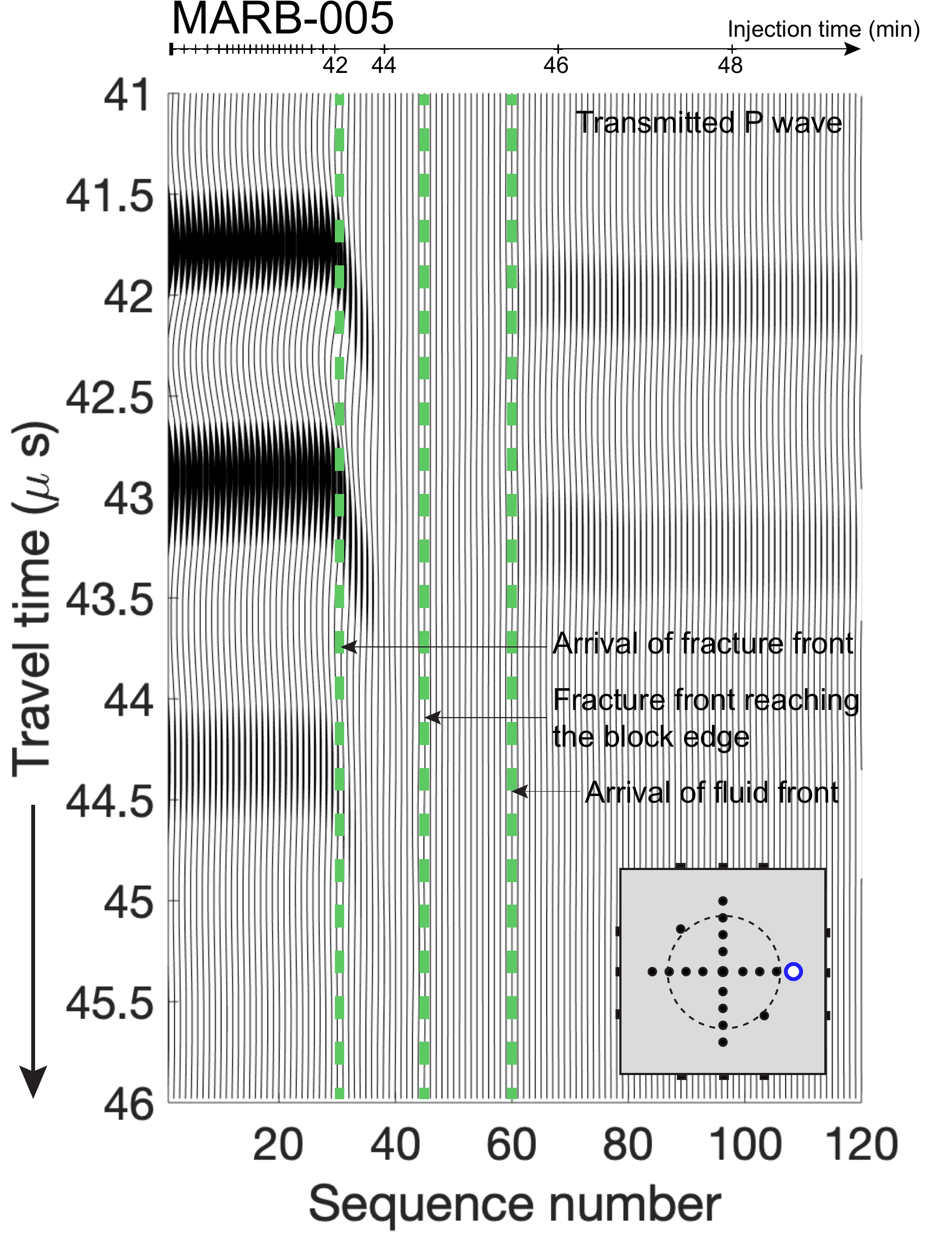}
    \includegraphics[width=0.48\linewidth]{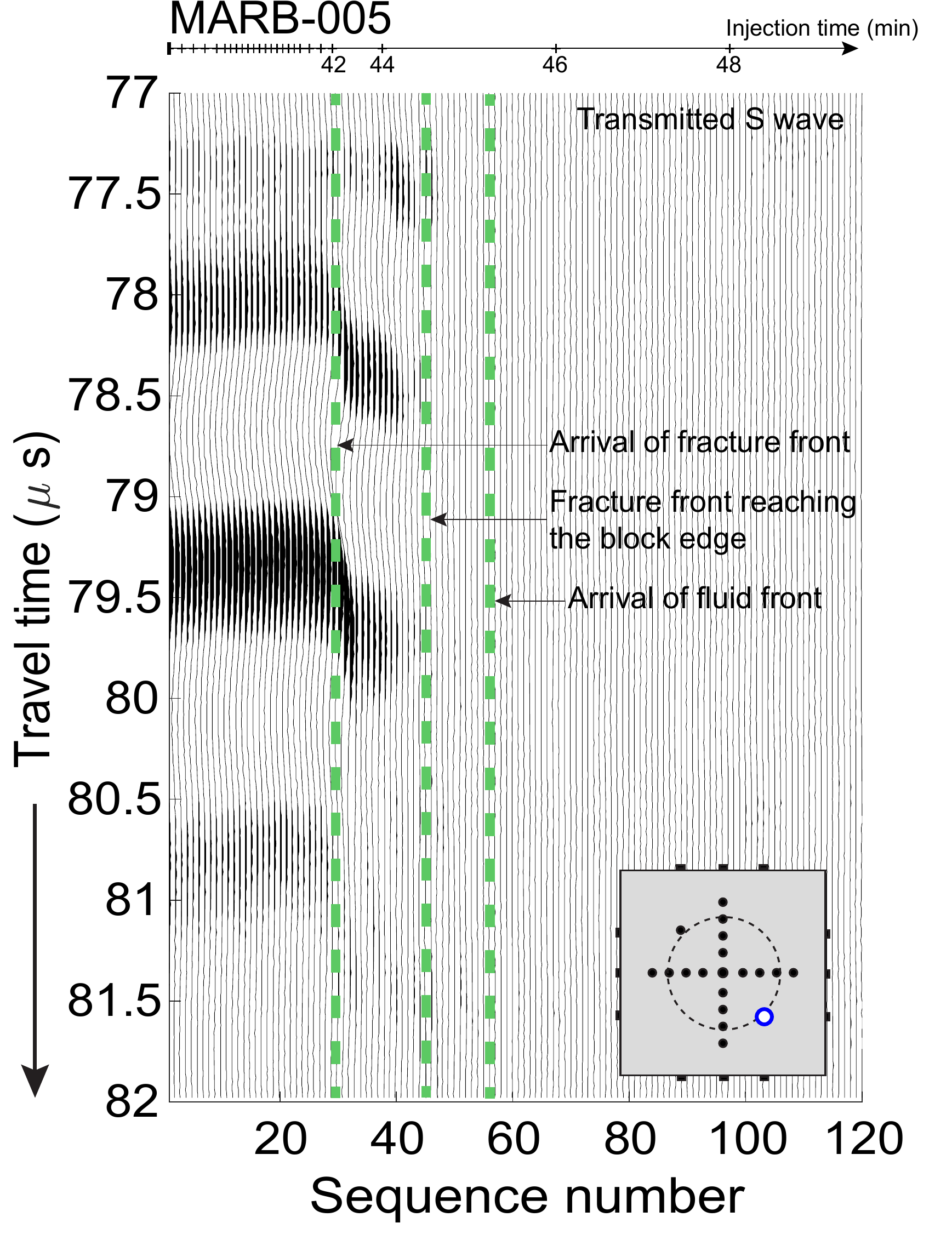}
    \caption{Evolution of the recorded signal for opposite source-receiver pair during the MARB-005 experiment: attenuation of the compressional (left) and shear wave (right).}
    \label{fig:MShearShadow}
\end{figure}
As illustrated in Fig.~\ref{fig:MShearShadow}, 
both compressional and shear waves significantly lose their amplitude
in between the time when the fracture and fluid front passes through the corresponding ray path.  Following the arrival of the fluid front, the compressional wave regains its amplitude but not the shear wave. Such a characteristic shear wave shadowing due to the lack of shear stiffness of the fluid 
is consistent with previous observations \citep{groenenboom2000monitoring}.

\begin{figure}
    \centering
    \includegraphics[width=0.9\linewidth]{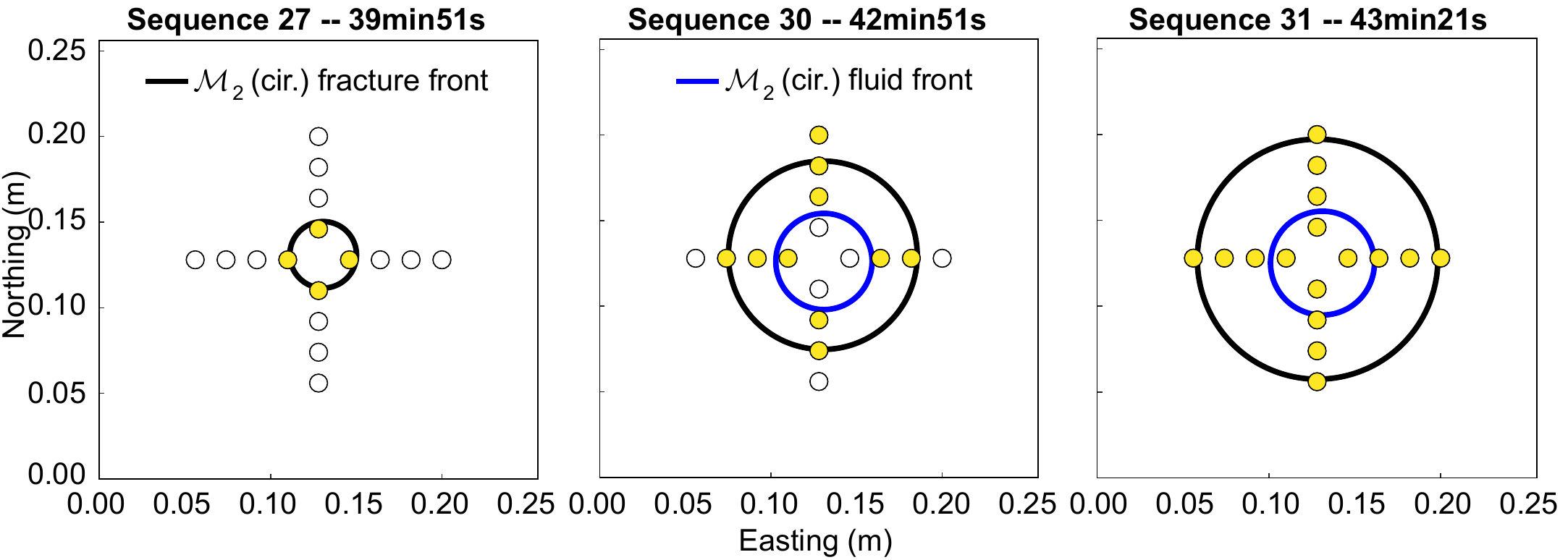}
    \caption{MARB-005 experiment: top view of the extent of the hydraulic fracture from transmitted waves (from top-bottom platens). The P-wave transducers turn yellow if the attenuation ratio of the signal strength $(E/E_{ref})^{1/2}$ goes below 0.8. We take sequence 16 (before fracture initiation) as the reference sequence for the calculation of the attenuation ratio. The corresponding footprint of the fracture and fluid fronts obtained from the inversion of diffracted waves are displayed for comparisons.}
    \label{fig:Mtransmission}
\end{figure}

We use the time evolution of the transmitted energy ratio defined in Eq.~(\ref{eq:transmissionenergy}) for all the pairs of compressional waves transducers in the opposite top and bottom platens (sub-parallel to the created fractures) and compare it with the fracture and fluid front previously reconstructed from diffracted waves (see Fig.~\ref{fig:Mtransmission}).
We use a threshold of 0.8 for $(E/E_{ref})^{1/2}$ to binarize the loss (for value below the threshold) of the transmitted wave. 

As shown in Fig.~\ref{fig:Mtransmission},  we first clearly see that the transmitted signal is lost when the reconstructed fracture front reaches the transducers location (black curve in the snapshot of Fig.~\ref{fig:Mtransmission}). The signal is then regained upon the arrival of the fluid front  (blue curve in the snapshot of Fig.~\ref{fig:Mtransmission}), but eventually lost again due to the increase of the fracture width as the viscous fluid front penetrates more into the fracture.
Overall, we observe in Fig.~\ref{fig:Mtransmission} a good agreement between the evolution 
of the signal strength ratio of the transmitted P waves with the evolution of the fracture and fluid fronts reconstructed from diffraction data. 

\subsubsection{Toughness-dominated experiment GABB-001}
\begin{figure}
    \centering
    \includegraphics[width=0.48\linewidth]{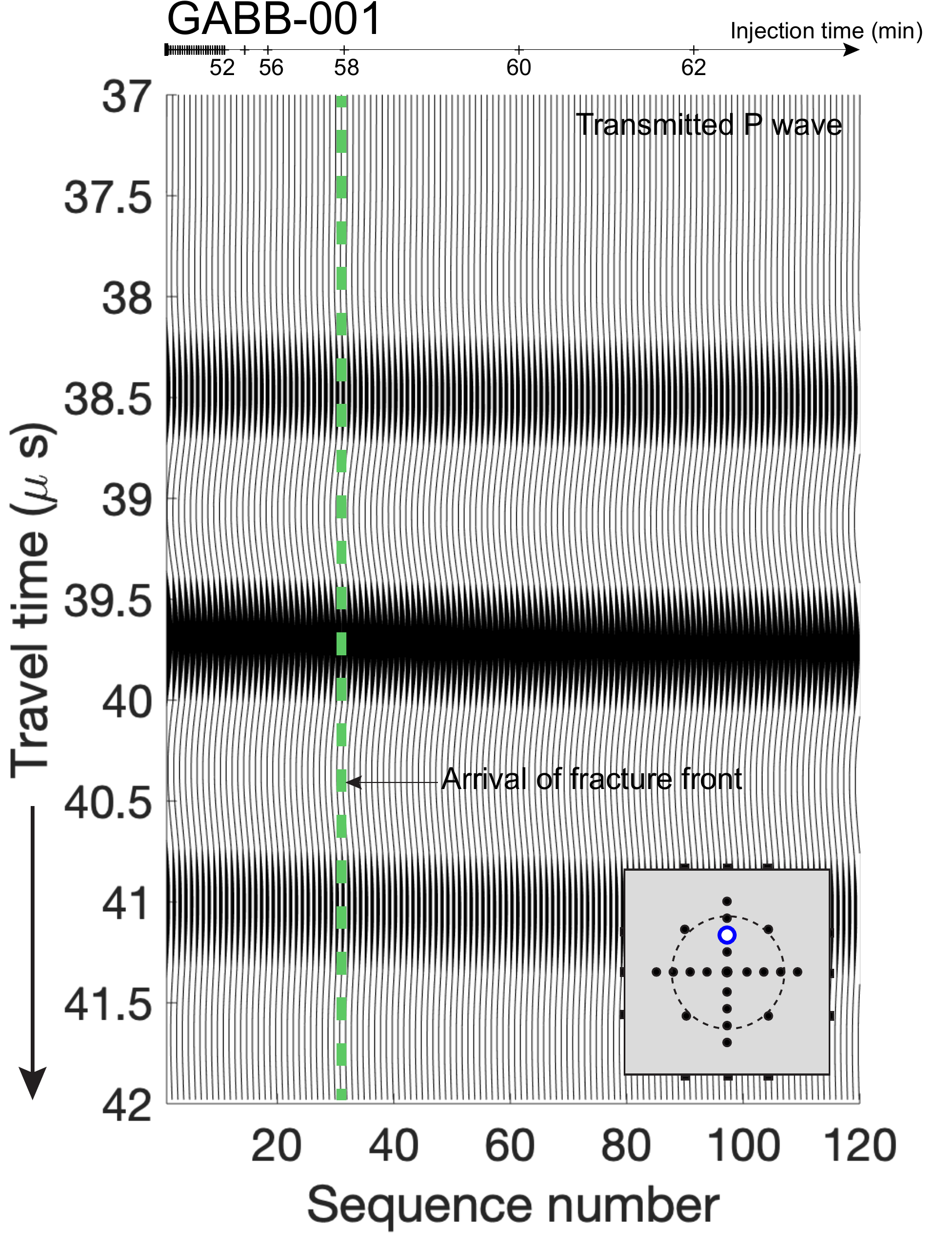}
    \includegraphics[width=0.48\linewidth]{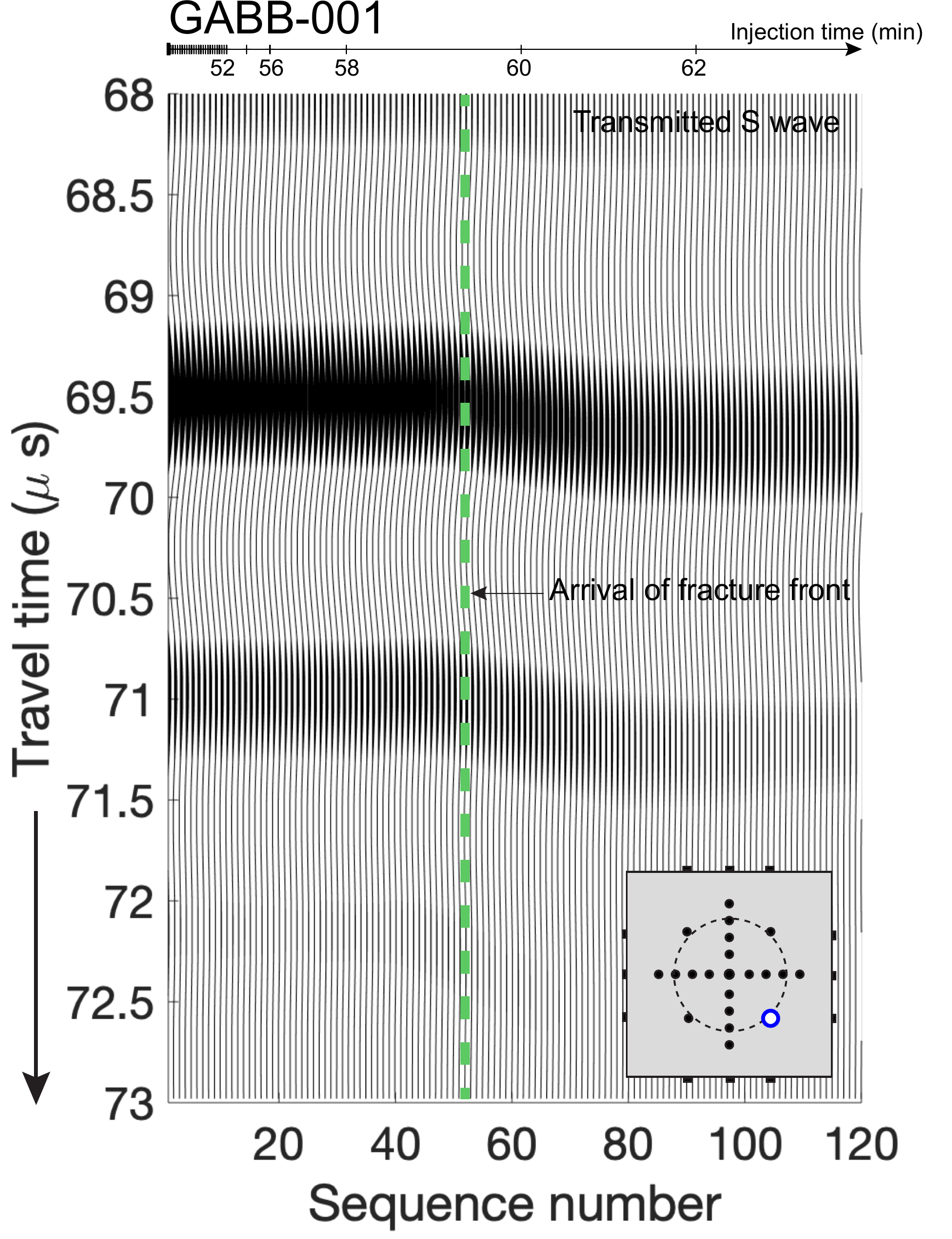}
    \caption{GABB-001 experiment: gradual attenuation of the compressional wave (left) and shear wave (right).}
    \label{fig:GShearShadow}
\end{figure}
In the GABB-001 experiment, the fracture front coincides with the fluid front (no fluid lag).  The transmitted shear waves present a gradual attenuation after the arrival of the fracture front as shown in Fig~\ref{fig:GShearShadow}. Such a weak shear shadowing effect is probably due to the smaller width of the fracture as well as the possible existence of solid bridges between fractured surfaces.
In order to separate apart the created fracture surfaces, 
we had to hammer a sub-sampled part of the specimen in gabbro. On the other hand, it is worth noting that the marble block was already completely separated after the experiment and exhibited smoother surfaces.

\begin{figure}
    \centering
    \includegraphics[width=0.9\linewidth]{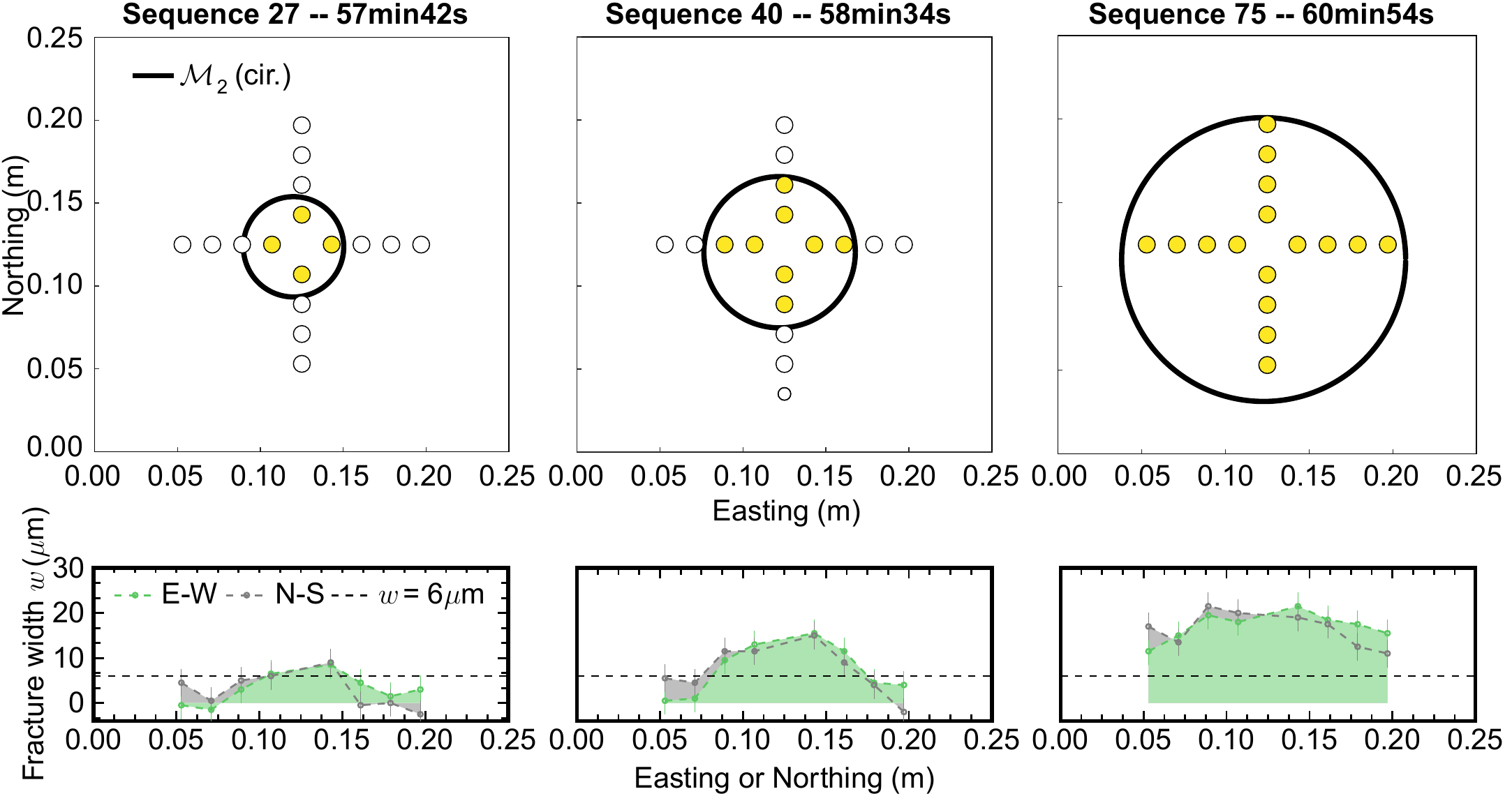}
    \caption{GABB-001 experiment: top view of the extent of the hydraulic fracture. The P-wave transducers turn yellow if the fracture opening goes above $6$~$\mu m$ with an error of around $3$~$\mu m$. This error on the fracture width estimation was obtained from the maximum width obtained when no fracture was present in the block (using acquisition sequences prior to fracture initiation).}
    \label{fig:Gtransmission}
\end{figure}

The compressional waves do not attenuate significantly during the fracture growth (Fig.~\ref{fig:GShearShadow}) but sufficiently to allow the reconstruction of the fracture width using the three layers model described previously.
The evolution of fracture extent grasped via the evolution of the fracture widths inverted for all the top bottom platens P transducers pairs is shown in Fig.~\ref{fig:Gtransmission}. It agrees relatively well with the fracture front reconstructed from the diffracted waves although a damage zone ahead of the reconstructed fracture tip may indeed exist. In addition,
the order of magnitude of the fracture widths is in line with the predictions of the toughness dominated solution for a radial hydraulic fracture \citep{savitski2002propagation}. By using the averaged entering flow rate $<Q_o>$ for estimation and the properties listed in Table~\ref{tab:ExpScaling}, the toughness dominated radial solution predicts respectively a maximum width (at the fracture center) of respectively 12~$\mu$m, 19~$\mu$m and 24~$\mu$m for the presented sequences.

\begin{figure}
    \centering
    \includegraphics[width=0.48\linewidth]{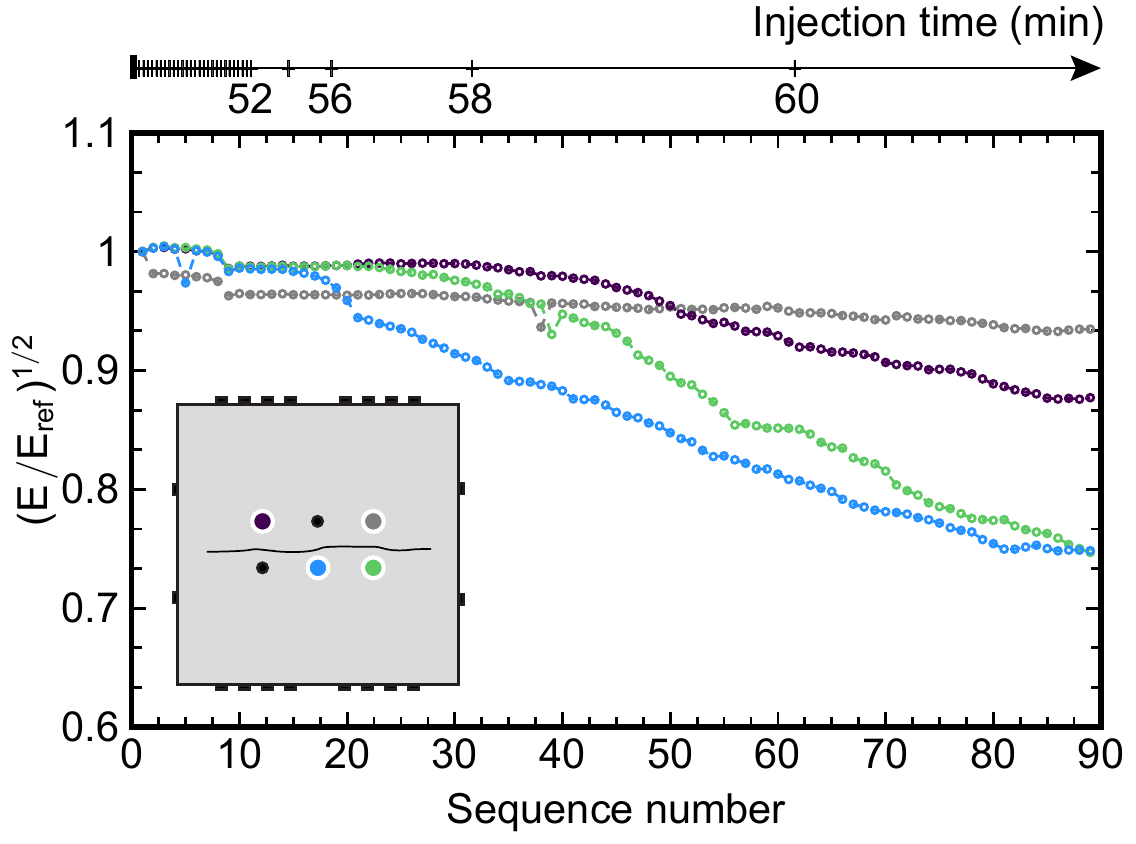}
    \caption{GABB-001 experiment: attenuation of the compressional waves propagating parallel to the fracture surface with a location of around 2 cm away from the fracture.}
    \label{fig:GSideTransmission}
\end{figure}

The existence of a "tappered" width profile near the tip associated with the existence of the process zone is not extremely clear (due to the resolution of the fracture width estimation). Another line of evidence for the presence of a process zone relates to attenuation of transmitted waves propagating parallel to the fracture plane (above and below it). 
We choose side transmitted pairs with a good signal to noise ratio and evaluate their attenuation ratio using Eq.~(\ref{eq:transmissionenergy}). Transducers on the north-southern sides of the block which are located approximately 5~cm away from the fracture plane present a difference of signal strength of around 5\% after fracturing. The transducer pairs on the west-eastern sides which are 2~cm from the fracture plane, present a larger attenuation as shown in Fig.~\ref{fig:GSideTransmission}. It looks like a band of  $\pm$  2~cm above and below the fracture plane is influenced.
Such an attenuation has two possible explanations. First, the waves interact with the presence of the fracture thus decreasing the received amplitude. Another possibility lies in the presence of micro-cracks surrounding the growing fracture which are known to strongly attenuate transmitted waves \citep{ZhGr93,zang2000fracture}.
Further analysis is required in order to decipher between these two explanations.

\section{Conclusions}

We have improved the resolution of the monitoring of growing hydraulic fracture using diffraction data recorded by a large amount of piezoelectric source/receiver pairs. Using Bayesian inversion, we have developed  a workflow to select the most probable fracture geometry from a finite number of simple models (circle, ellipse, horizontal or not). This model ranking allows to gain confidence in the important possible deviation of the fracture from the simplest radial shape.
Moreover, the inversion of the modelling error $\sigma$ (which combines model and measurement error) allows to quantify the ability of any chosen model to reproduce the data. 
The method has been successfully tested on two experiments representative of two very different HF propagating regimes (toughness and lag-viscous dominated). In both cases, the fractures were mostly radial although a deviation towards a more elliptical shape is visible when the fracture feels the edge of the specimen (where the applied stress field is likely less uniform). 
Although it is difficult to precisely gauge the accuracy of the reconstruction, 
the resulting posterior uncertainties of the fracture extent are around 1-2~mm for the gabbro experiment (GABB-001) and 2-4~mm for the marble experiment (MARB-005). 
The fronts reconstructed from diffracted waves agree well with the analysis of the attenuation of compressional waves traversing the propagating fractures. 
It is also important to recall that in our analysis, the data for one acquisition sequence is assumed to be acquired at the same time while an acquisition lasts about 2.5~seconds (spanning of all the sources). As a result, this imaging technique is appropriate only for low velocity fractures: the average fracture velocity is around 300 $\mu$m/s  for GABB-001, and 1~mm/s for the (faster) MARB-005 experiment.

The method presented here can be improved in a number of ways. First, instead of using parametrized fracture shapes, a direct extension is to use a 3D spline curve to describe the fracture front (at the expense of more model parameters). Secondly, in order to better quantify the possible damage around the growing fracture, the hypothesis of  a constant wave velocity in the sample during fracture growth should be at least partly relaxed (to account for the effect of possible micro-cracking around the fracture). 
It would be interesting to combine the analysis of diffracted waves with recent acoustic tomography inversion that reconstructs such velocity changes in the bulk
using only direct wave arrivals but combining passive and active acoustic data  \citep{brantut2018time,aben2019rupture}. 
Full waveform inversion (likely in the frequency domain) would ultimately allow to combine the information of diffracted, transmitted (and reflected) waves, but this requires proper sensor calibration and the use of  computationally expensive models able to resolve the sharp discontinuities induced by the fluid-filled fracture. 
A more immediate/simpler improvement will likely come from the combination of the active 4D acoustic method presented here with passive listening for AE events (with localization and moment tensor estimation, see for example \cite{StBu15}). This will surely enhance our understanding of hydraulic fracture growth, in particular with respect to a better quantification of the fracture process zone in rocks. 
In particular, the interplay between the evolution of the fluid lag and the process zone appears to be strongly influencing the overall HF propagation according to recent theoretical predictions \citep{garagash2019cohesive}.

\paragraph*{Acknowledgements\\}
{
Partial support from the SCCER-SoE (second phase 2017-2020) funded by the Swiss National Science Foundation and the Swiss Innovation Agency InnoSuisse is greatly acknowledged.
}

\paragraph*{Data availability \\}
{
The raw and processed data  of these two experiments as well as the inversion results will be made available via the Zenodo platform.
}

\paragraph*{CRediT Authors contributions \\}
{
 Dong Liu: Conceptualization, Methodology, Data curation, Formal analysis, Visualization, Writing – original draft. \\
Brice Lecampion: Conceptualization, Methodology, Formal analysis, Supervision, Resources, Writing – review \& editing. \\
Thomas Bum: Conceptualization, Methodology, Data curation, Formal analysis.
}

\appendix
\section{Evolution of the entering flow rate into the fracture}\label{app:inletflux}

 The interface vessel introduces a non-negligible system compliance. As a result, the flow rate entering into the fracture $Q_{in}$ does not  equal the pump injection rate $Q_o$ upon fracture initiation. One can estimate $Q_{in}$ based on the fluid pressure measurement and the estimation of the system compliance, using the global mass balance of fluid in the injection line from the pump to the fracture inlet. We thus obtain
\begin{equation}
    Q_{in}(t)=Q_o-c_f V_d \frac{\text{d}p_{\text{downstream}}(t)}{\text{d}t}-(U-c_fV_d) \frac{\text{d}p_{\text{upstream}}(t)}{\text{d}t}
    \label{eq:Qin}
\end{equation}
where $p_{\text{upstream}}$ and $p_{\text{downstream}}$ represent respectively the fluid pressure upstream and downstream as shown in Fig.~\ref{fig:Transducers}, and $c_f$ the fluid compressibility and $V_d$ the fluid volume downstream (from needle valve to fracture notch). The system compliance $U$ can be estimated from the averaged pressurization rate before fracture initiation.
\begin{equation}
    U=Q_o/\left(\frac{\text{d}p_{\text{upstream}}}{\text{d}t}\right)
\end{equation}


\bibliographystyle{gji}

\label{lastpage}
\end{document}